\documentclass[final,5p,times,twocolumn,authoryear]{elsarticle}

\usepackage{graphics}
\usepackage{natbib}
\usepackage{multirow}
\usepackage{rotating}
\usepackage{amsfonts}
\usepackage{amsmath}
\usepackage{enumerate}
\usepackage{tabularx}
\usepackage{natbib,amsmath,amssymb,epsfig,graphicx,lscape,longtable,multirow,float,url,footnote}
\usepackage{lineno}
\usepackage{amssymb}
\usepackage{morefloats}

\usepackage{url}

\newcommand{\degr} {$^{\circ}$}

\newcommand{\beq}{\bigskip\begin{equation}}
\newcommand{\eeq}{\bigskip\end{equation}}
\newcommand{\prs}{$^{\prime\prime}\;$}
\newcommand{\pr}{$^{\prime\prime}$}

\begin{document}

\setcounter{table}{0}
\setcounter{figure}{0}

\begin{frontmatter}

\title{Jupiter's Ammonia Distribution Derived from VLA Maps at 3--37 GHz 
}




\author[idp1,idp7]{Imke de Pater}
\author[idp2]{R. J. Sault}
\author[idp1]{Michael H. Wong}
\author[idp3]{Leigh N. Fletcher}
\author[idp1]{David DeBoer}
\author[idp4]{Bryan Butler}

\address[idp1]{Astronomy Department, 501 Campbell Hall, University of California, Berkeley, CA 94720}
\address[idp2]{School of Physics, University of Melbourne, Victoria, 3010, Australia}
\address[idp7]{Faculty of Aerospace Engineering, Delft University of Technology, 2629 HS Delft, The Netherlands}
\address[idp3]{Department of Physics and Astronomy, University of Leicester University Road, Leicester, LE1 7RH, UK.}
\address[idp4]{National Radio Astronomy Observatory, Socorro, NM 87801, USA}

\begin{abstract}
We observed Jupiter four times over a full rotation (10 hrs) with the upgraded Karl G. Jansky Very Large Array (VLA) between December 2013 and  December 2014. Preliminary results at 4-17 GHz were presented in de Pater et al. (2016); in the present paper we present the full data set at frequencies between 3 and 37 GHz.  Major findings are:

{\it i)} The radio-hot belt at 8.5--11$^\circ$N latitude, near the interface between the North Equatorial Belt (NEB) and the Equatorial Zone (EZ) is prominent at all frequencies (3--37 GHz). Its location coincides with the southern latitudes of the NEB (7--17$^{\circ}$ N).

{\it ii)} Longitude-smeared maps reveal belts and zones at all frequencies at latitudes $\lesssim |20^\circ|$. At higher latitudes numerous fainter bands are visible at frequencies $\gtrsim$ 7 GHz. The lowest brightness temperature is in the EZ near a latitude of 4$^\circ$N, and the NEB has the highest brightness temperature near 11$^\circ$N. The bright part of the NEB increases in latitudinal extent (spreads towards the north) with deceasing frequency, i.e., with depth into the atmosphere. In longitude-resolved maps, several belts, in particular in the southern hemisphere, are not continuous along the latitude line, but broken into small segments as if caused by an underlying wave. 

{\it iii)} Model fits to longitude-smeared spectra are obtained at each latitude. These show a high NH$_3$ abundance (volume mixing ratio $\sim 4 \times 10^{-4}$) in the deep ($P>8-10$ bar) atmosphere, decreasing at higher altitudes due to cloud formation (e.g., in zones), or dynamics in combination with cloud condensation (belts). 
In the NEB ammonia gas is depleted down to at least the 20 bar level with an abundance of $1.75 \times 10^{-4}$. 
The NH$_3$ abundance at latitudes $>|50|^\circ$ is characterized by a relatively low value ($\sim 1.75 \times 10^{-4}$) between $\sim$~1 and 10 bar. 

 {\it iv)} Using the entire VLA dataset, we confirm that the planet is extremely dynamic in the upper layers of the atmosphere, at $P<$2--3 bar, i.e., at the altitudes where clouds form. At most latitudes the relative humidity within and above the NH$_3$ cloud is considerably sub-saturated. 

{\it v)} The radiative transfer models that best fit the longitude-smeared VLA data at 4--25 GHz match the {\it Juno} PeriJove 1 microwave data extremely well, i.e., the NH$_3$ abundance is high in the deep atmosphere, and either remains constant or decreases with altitude.

{\it vi)} Hot spots have a very low, sub-saturated NH$_3$ abundance at the altitudes of the NH$_3$-ice cloud, gradually increasing from an abundance of $\sim 10^{-5}$ at 0.6 bar to the deep atmosphere value ($\sim 4 \times 10^{-4}$) at 8 bar.

{\it vii)} We previously showed the presence of large ammonia plumes, which together with the 5-$\mu$m hot spots constitute the equatorially trapped Rossby wave. Observations of these plumes at 12--25 GHz reveal them to be supersaturated at $\sim$~0.8--0.5 bar, which implies plumes rise $\sim$~10 km above the main clouddeck. Numerous small ammonia plumes are detected at other locations (e.g., at 19$^{\circ}$S and interspersed with hot spots).

{\it viii)} The Great Red Spot (GRS) and Oval BA show relatively low NH$_3$ abundances throughout the troposphere ($\sim$~1.5--1.8 $\times$ 10$^{-4}$), and the GRS is considerably sub-saturated at higher altitudes.

\end{abstract}


%


\end{frontmatter}





\section{Introduction}

A few years after the serendipitous detection of Jupiter's strong radio bursts near 22 MHz, known as the planet's decametric radiation (Burke and Franklin, 1955), Mayer et al. (1958) reported an observation of the planet's microwave radiation at 3 cm wavelength, which constituted the first detection of the planet's thermal emission at radio wavelengths. Subsequent observations at longer wavelengths soon revealed a third component of Jupiter's radio emissions:  synchrotron radiation emitted by high energy electrons trapped in Van Allen-like radiation belts (Radhakrishnan and Roberts, 1960). The synchrotron emission dominates at wavelengths $\gtrsim$~6 cm, and separation of the thermal and nonthermal components is necessary to analyze both types of emissions. This separation has been accomplished via measurements of the polarization, and/or by separating the two components spatially; it does remain tricky, though (e.g., Berge and Gulkis, 1976; de Pater et al., 1982). 

The first spatially-resolved observations of Jupiter's disk were obtained with the Very Large Array soon after the array was commissioned (de Pater and Dickel, 1986). The images show the familar zone-belt structure. In particular, at a wavelength of 2 cm the North Equatorial Belt (NEB) was prominent. It was  $\sim$~15K warmer than the Equatorial Zone (EZ), while the South Equatorial and South Tropical Belts (SEB, STB) as well as both polar regions were $\sim$~5--8 K warmer than predicted based upon radiative transfer calculations for a solar composition atmosphere and an adiabatic temperature-pressure (TP) profile. The brightness temperature variations were interpreted by spatial variations in the ammonia abundance, the main source of opacity at cm wavelengths (de Pater, 1986). 

The first longitude-resolved map was published in 2004 (Sault et al., 2004). This map revealed for the first time radio hot spots which appeared to be co-located with the 5-$\mu$m hot spots. The ammonia abundance, defined in this paper as the volume mixing ratio, in these hot spots was estimated at about half the abundance of that in the NEB,  $\sim$~3$\times$10$^{-5}$, down to $P>$ 4 bar. 

In the present paper we present both longitude-smeared (zone-belt structure) and longitude-resolved maps of an extensive dataset obtained with the upgraded Karl G. Jansky Very Large Array (VLA) in 2013-2014, at $\sim$~3--37 GHz. A subset of these data (4-17 GHz) was published before (de Pater et al., 2016; hereafter referred to as dP16), and revealed a radio-hot belt just north of the eastward jet at 7.8$^\circ$N, at the interface between the NEB and EZ, dotted with longitudinally-periodic hot spots. This radio-hot belt extends over the latitude range 8.5--11$^\circ$N, i.e., it is located on the south side of the NEB latitude range (6.9--17.4$^{\circ}$ N; Fletcher et al., 2016).  Just south of this radio-hot belt, periodic radio-cold features at latitudes of 2--5$^\circ$N revealed ammonia-rich plumes. We interpreted these hot spots and plumes as markers of the same equatorially trapped Rossby wave theorized to create the 5-$\mu$m hot spots seen in infrared observations (Showman and Dowling, 2000; Friedson, 2005). 

Numerous vortices and small scale structures in these maps revealed a planet that is extremely dynamic at pressures less than 2--3 bar, i.e., in Jupiter's ``weather'' layer within and above the NH$_4$SH cloud layer. At the deeper levels probed at 4-7 GHz frequencies, these small features were no longer seen, even though the radio-hot belt, hot spots, and plumes at $\sim$~2--12$^\circ$N remained prominent. These lower-frequency data showed that the low ammonia abundance in hot spots extends down to at least the 8-bar level, and that ammonia gas in the radio-cold plumes is brought up from the deep atmosphere at the abundance prevalent at those deep layers, i.e., close to the abundance derived from the Galileo probe data by Wong et al. (2004a).

The microwave experiment on the {\it Juno} spacecraft showed that the NEB, which includes the radio-hot belt on its south-side, extends down to many tens of bars (Bolton et al., 2017; Li et al., 2017).

In Section 2 we discuss the observations and data reduction techniques to create longitude-smeared and longitude-resolved maps in detail; special attention is given to determining absolute brightness temperatures on these maps. In Section 3 we discuss our radiative transfer (RT) code, Radio-BEAR, which is used in subsequent sections to interpret the data. Section 4 shows results, starting with disk-averaged spectra, then longitude-smeared and longitude-resolved maps. A detailed analysis with a discussion is provided in Section 5; this section also contains a comparison to {\it Juno}/MWR and {\it Cassini}/CIRS observations. The final section provides a summary of our findings.

\section{Observations and Data Reduction}\label{obs}

\subsection{General}\label{obs1}

Radio observations of Jupiter were conducted with the upgraded Very
Large Array (VLA) (Perley et al., 2011). We observed Jupiter for 10
hrs on each of 4 days, each day with a different array
configuration. We generally observed each frequency band with two
array configurations: a low and a high resolution configuration.  In
the A configuration, the antennas have a maximum separation of 36 km,
with a minimum spacing of 0.68 km. In contrast, in the most compact or
D configuration, the longest spacing between antennas is 1 km, and the
shortest spacing is 0.035 km. The B configuration has spacings between
0.21 km and 11.1 km, and the C configuration between 0.035 km and 3.4
km. By carefully choosing our configurations and frequencies we were
able to cover Jupiter at all frequencies between 1 and 37 GHz with
both long and short spacings (high [$\lesssim$~0.6\pr] and low [$\gtrsim$~4\pr] resolution, respectively),
as summarized in Table 1. This table provides also the frequencies/wavelengths associated with the various observing bands (i.e., K, Ku, Ka, C, S, L).

The basic data received from an interferometer array, such as the VLA,
are (complex) visibilities, formed by correlating signals from the
array's elements. These are measured in the u-v plane, where the
coordinates u and v describe the separation, or baseline, between two
antennas (i.e., an interferometer) in wavelength, as projected on the
sky in the direction of the source. For short baselines, the u-v
points lie close to the center of the plane; for long baselines, they
are at larger distances. An image of the source is then obtained by
Fourier transforming the visibilities. Since the u-v plane is only
sampled at particular spacings, depending on baseline and wavelength,
an interferometer array spatially filters its observations.  Hence, as
well as being insensitive to features finer than a particular
resolution, radio interferometer arrays are also insensitive to
features that are broader than a particular scale-size, i.e., images
produced by radio interferometer arrays spatially filter out the low
spatial frequency (i.e. broad scale) structure. This is unlike many
other sorts of telescopes. This spatial filtering is determined by the
shortest physical separation of a pair of antennas in the array.

For a source such as Jupiter, which has complex structure on all
spatial frequencies, this creates a challenge. This is more
challenging still because the high spatial frequency features are time
varying. We have attempted to address these challenges by observing in
two array configurations at most frequencies. The predictable changes
caused by variations in the distance between Jupiter and the observer
have been normalized out; all data were normalized to the 23 December
2013 date. However, our various steps have been only partially
successful at addressing the challenges of the spatial filtering. In
the analysis that follows, a number of checks and ad hoc steps have
been used to further ameliorate the spatial-filtering effects that are
inherent in the observations.

The flux density scale was calibrated on the radio source 3C286, for which we used the standard
VLA flux calibrator scale (Perley and Butler, 2013, 2017).  Internal
and absolute uncertainties in the flux densities are believed to be
better than $\sim$~3\%, perhaps rising to 5\% at the highest
frequencies. We adopt a 3\% calibration error at all frequencies. However, as shown in Section~\ref{spec}, this is a rather conservative estimate.

The initial processing of the data, such as flagging and editing (e.g., deleting bad data)
was done via the internal VLA
calibration pipeline. After careful inspection of the data, we noticed
that the flux density scale as derived and applied by the pipeline was
inconsistent between different datasets, as initially pointed out in dP16. 
All data, therefore, were
recalibrated by hand, using AIPS (NRAO; http://www.aips.nrao.edu/CookHTML/CookBook.html) and the MIRIAD software package
(Sault et al., 1995).  Each dataset was also inspected for interfering
background sources; several background sources were subtracted in the
L-band B-configuration dataset. Two sources that did interfere with Jupiter in several of the higher-frequency data sets were the
satellites Io and Europa. By phase shifting the u-v data such as to
follow the motion of these satellites, their contribution was
subtracted from the u-v data. Afterwards the u-v data were phase
shifted back to follow the motion of Jupiter on the sky. All data
sets between 4 and 37 GHz were then split into roughly 1-GHz wide chunks (Table 2), each
of which was mapped separately. At S (3 GHz) and L (1.4 GHz) bands we combined all the data per band (i.e., resulting in one S band and one L band image).

Each of these data sets was self-calibrated. Self-calibration uses a model
of the visibilities to derive antenna-based corrections to the
visibilities so that the visibilities over time are self-consistent in
the end (e.g., Butler et al.  2001). Our model was the limb-darkened disk that best matched the observations, sometimes (e.g., for X band)
augmented with bands to represent the zone-belt structure. Although in principle both the amplitude and
phase can be corrected, we only corrected the phase of the
visibilities, since corrections to the gain can have disastrous consequences (such as obliterating bands, or rings in the saturnian system). 

In order to best assess small variations on Jupiter's disk, a
limb-darkened disk was subtracted from the u-v data with a brightness
temperature and limb-darkening parameter that produced a best fit (by eye) to
the data. Limb-darkening was modeled by multiplying the brightness
temperature $T_b$ by (cos$\theta$)$^q$, with $\theta$ the emission
angle on the disk (i.e., the angle between the surface normal vector and the line-of-sight vector to Earth), and $q$ a constant that provides a best fit to the
data. Both $T_b$ and $q$ varied with frequency to give a most accurate
representation of the disk as a function of frequency (Table 2). Since we had
split the data into 1-GHz wide chunks, we could match the data
reasonably well with these parameters, though not perfectly. In C and X band 
a bright ellipse was left on the limb of the planet after
subtraction of a limb-darkened disk. To minimize this effect, we
scaled the size of the planet that was subtracted by the small size-scaling
factor listed in Table 2. Perhaps if we had used a much more complex limb-darkening model, and/or had split the data in much finer frequency bins (e.g., 50--100 MHz), this bright ellipse would not have been present. However, our simple size-scaling solution was sufficient and easy to implement, and kept the radiative transfer analysis relatively straightforward. 

\subsection{Longitude-smeared Maps and Disk-averaged Brightness Temperatures}\label{smeared1}

Longitude-smeared maps of each full band (Ka, K, Ku, X, C, S and L),
as well as of each 1-GHz wide dataset were produced. Although the NEB
could be distinguished in the L band map, the thermal emission was
contaminated too much by Jupiter's synchrotron radiation to be
useful, and hence we will not discuss the L band data further in this
paper. The data from the high resolution configurations, i.e., the X
and Ku band in the B configuration, C band in the A configuration, and
K band in the C configuration, best show the latitudinal variations
across the disk. These maps, after subtracting off a limb-darkened
disk as described in Section~\ref{obs1}, are displayed in
Figure~\ref{fig1}. To account for the changing distance between
Jupiter and the observer, all data were normalized to the epoch of 23
December 2013, and all maps were convolved down to the same angular
resolution of 0.8'', except for the Ka band data, which have a
resolution of 2.5''. Numerous zones and belts are visible at K, Ku and X bands, while the NEB, SEB and EZ are
prominent even at C and S bands.

These maps, however, may suffer from a lack of short spacing data, visible as a
negative bowl underlying the disk (see e.g., de Pater et al., 2001); such a bowl can be discerned in the X-band map.
Such missing short spacings make it difficult to determine the
absolute brightness temperature. Unfortunately, the data from the
lower resolution configurations have too poor a spatial resolution to
be useful for any analysis of brightness variations across the
disk. However, these data have been used to help determine the disk-averaged brightness temperatures (Section~\ref{Tbav}), 
and to assess the accuracy in brightness temperature values for spatial variations across the disk (Section~\ref{Tbspatial}).

\subsubsection{Disk-averaged Brightness Temperatures}\label{Tbav}

We determined the disk-averaged brightness temperatures reported in
the last column of Table 2 as follows:

{\it K band (22 GHz, 1.4 cm)}: We made maps of the combined D$+$C (i.e., low$+$high resolution) configurations,
after subtraction (in the u-v plane) of a limb-darkened disk
with the parameters as listed in columns 4--6 in Table 2. For the low resolution configuration we only used data at
spacings $<$ 12 k$\lambda$ (1 k$\lambda$ $=$ 1 kilo-wavelength) to avoid
problems due to time variations and/or changes in viewing
geometry. Column 7 provides the disk-averaged brightness temperature
for the disk that was subtracted (T$_b$(av)), column 8 shows the
residual disk-averaged brightness temperature (T$_{\rm res}$), column
9 the brightness temperature that corresponds to the cosmic background
radiation (cmb) at that frequency (T$_{\rm cmb}$; e.g., Gibson et al.,
2005). Column 10 reports the expected contamination of the synchrotron
radiation as a disk-averaged brightness temperature (T$_{\rm
  synch}$). This number, 6--7\% of the total synchrotron radiation at
that frequency, was estimated from spatially-resolved models of
Jupiter's synchrotron radiation (de Pater et al., 1997). By using
spectra of the total synchrotron radiation (de Pater et al., 2003; de
Pater and Dunn, 2003), we determined the contaminating flux density at
each fequency. The final disk-averaged brightness
temperatures:
T$_b$(map)=T$_b$(av)+T$_{\rm res}$+T$_{\rm cmb}$-T$_{\rm synch}$ are listed in Column 11. 

We
independently determined the total flux density by fitting a disk to all
the u-v data of the low resolution (D for K band) configuration; this number,
T$_b$(UV), after subtraction of the synchrotron radiation (total value, $\sim$~1 Jy at $\Delta =$ 4.04 AU --  see de Pater and Dunn, 2003) and addition
of T$_{\rm cmb}$, is listed in Column 12. We averaged the 
numbers in Columns 11 and 12 to obtain our final disk-averaged temperature, T$_b$(final),
listed in Column 13. As stated in Section~\ref{obs1}, we adopted an
uncertainty of 3\% (adding, in quadrature, uncertainties in T$_{\rm synch}$ or any of the other 
components in T, does not noticeably affect the error). 

{\it Ku band (15 GHz, 2 cm)}: We made maps of the combined B$+$C configurations,
where the B configuration for January 2014 and December 2013 were
averaged, and only C configuration data at $<$ 12 k$\lambda$ were
used. The final disk-averaged brightness temperature (Column 13) was
determined in the same way as for the K band data, with T$_b$(UV)
determined from the low resolution (C) configuration data.

{\it C band (6 GHz, 5 cm)}: In contrast to the K and Ku bands, where the synchrotron
radiation is a very small fraction of the total radiation, at C
band the total synchrotron radiation is comparable in flux density to the total
thermal radiation, and extends out to almost three jovian radii from
the center of the planet. 

To understand our approach to handling this it is necessary to
appreciate that a radio interferometer array measures a spatially
filtered version of the source. For the high resolution (larger array)
configurations, the broader spatial structure is filtered out by the
interferometer array. For higher resolution configurations, the detected
thermal emission is dominated by the sharp transition at the edge of the
disk, as well as features on the planetary disk. However, because the
synchrotron emission is much smoother, it can be filtered out (``resolved out'')
by the interferometer. That is, the interferometer can be insensitive to
the synchrotron emission. 
 
For C band, in the A configuration the synchrotron radiation is mostly
resolved out, whereas in the C configuration most of the synchrotron
emission is partially (but poorly) sampled. We found we would get
unreliable measures when we attempted to estimate total intensity levels
using both array configurations. By using only the high resolution (A)
configuration data and subtracting a best-fit limb-darkened disk from
the u-v data, much of the contamination problem was avoided. In this high resolution configuration, in both total intensity and linearly polarized maps only emission at the peak of the radiation belts was visible, without a trace on the disk (also after taking out the ``wobble'' in synchrotron radiation due to the $\sim$~10$^\circ$ tilt of the magnetic axis relative to the rotation axis, so as to obtain the sharpest possible images of the synchrotron radiation).  
We therefore
determined the final disk-averaged brightness temperature from just the
high resolution data: $T_b(map)=T_b(av)+T_{\rm res}+T_{\rm cmb}$. Due to
the high level of synchrotron radiation in the low resolution data we
were unable to fit a disk to the data, so Column 12 is empty. As
discussed below (Fig. 4), the derived values for the disk-averaged
brightness temperatures agree well with previous measurements.

{\it X band (10 GHz, 3 cm)}: Relative to the thermal emission, the synchrotron radiation at X band is less pronounced than at C band, but much more than at Ku band. We determined the disk-averaged brightness temperature from the high resolution data only, using the combination of the December 2013 and January 2014 data. These values are listed in Column 11. However, when plotted as a spectrum we noticed that these numbers were 3\% too low (Fig.~\ref{figspec} discussed below), likely caused by a lack of short spacing data; in Column 13 we list the final values, where we had multiplied the temperatures in Column 11 by a factor of 1.03. 

{\it S band (3 GHz, 10 cm)}: At S band the synchrotron radiation overwhelms the thermal emission. We did determine the total emission from the A configuration data only, but it clearly was much lower than expected, $\sim$~240 K versus an expected $\sim$~270 K. We therefore multiplied the S band data by a factor of 1.13, but do not display this data point in the disk-averaged spectra discussed below. 

{\it Ka band (33 GHz, 0.9 cm)}: Ka band was observed in the D configuration only. Although this is the most compact configuration at the VLA, it is missing the shortest spacings at these frequencies. We determined the disk-averaged brightness temperature as for the C and X bands: T$_b$(map)=T$_b$(av)+T$_{\rm res}$+T$_{\rm cmb}$. As shown below, these numbers agree very well with brightness temperatures derived from a very compact interferometer array (Karim et al., 2018). 

\subsubsection{Laitudinal Variations in Brightness Temperature}\label{Tbspatial}

The maps we created from the high resolution configurations alone, after subtraction of a limb-darkened disk (Fig.~\ref{fig1}) look very good, which suggests that subtraction of such a disk did help mitigate the lack of the shortest spacings to a large extent. In order to assess the accuracy of spatial variations in the brightness temperature on such maps we conducted two experiments. 

$1)$ The first test compared contrasts in a known simulated disk (low Tb in the EZ vs high Tb in the NEB) to contrasts measured after processing the simulated visibilities identically to the observed visibilities. In detail: We created a uniform limb-darkened disk the size of Jupiter, with the parameters appropriate for the particular frequency (Table 2); we then superposed a --30 K band (to simulate the EZ), and +20 K band (the NEB) on the disk, and created simulated visibilities matching the characteristics of our observations. These artificial visibilities of the disk, with a u-v coverage that is the same as for the actual data, were then processed in the same way as the observations, and we evaluated the peak-to-peak (low T$_b$ in the EZ vs high T$_b$ in the NEB) difference on the maps. For the high resolution C band data this resulted in a contrast that was too low by 10--15\%; in Ku band it was $\sim$~5\% too low, and there was no loss in X, K, and S bands.

$2)$ The second test compared contrasts in high and low resolution maps. In detail: We compared the peak-to-peak (low EZ, high NEB) values on maps created from just the high resolution data with maps created from just the low resolution data, or data with a combination of configurations. These maps were all constructed with the same spatial resolution for both high and low resolution data. For example, for C band we used an angular resolution of 4.4\prs for both high and low resolution data. This process showed that the low EZ to high NEB brightness temperature contrast in the high resolution (A) configuration for C band was 30--60\% lower than in the low resolution (C) configuration (the range indicates the variation in values between the individual 1-GHz wide maps). 
In X band the temperature contrast in the high resolution (B) maps was 10--20\% lower than in the low resolution (B$+$C) maps, and in Ku band the difference in contrast was 15--30\%. At K band, the high resolution (C) configuration maps showed a contrast that varied from 0 up to 7\% less than in the low resolution (C$+$D) configuration maps.

Both tests showed that the brightness temperature contrast on the high resolution maps was usually less than at low resolution, though the difference in contrast was smaller in our simulated (noiseless) disk data. In order to at least partially correct for such contrasts, we multiplied our (disk-subtracted) maps by factors of 1.30 in C and S bands, 1.05 in X band, and 1.15 in Ku band. We did not apply a correction to the K band. In addition, at Ku and K bands we multiplied the maps (after adding the disks back) by the ratio of the disk-averaged brightness temperature listed in Column 13 in Table 2 and the disk-averaged brightness temperature derived from the high resolution data alone (this resulted in adjustments $<$3\%). As discussed in Section~\ref{smeared2}, uncertainties in any observed latitudinal contrast were increased in C and S bands to account for uncertainties in these factors.

\subsection{Longitude-resolved Maps}\label{resolved1}

The data reduction to obtain the longitude-smeared radio maps discussed above are what we refer to
as ``conventional radio interferometric images'', which usually are
integrated over many hours to meet the required sensitivity, while
Earth rotation synthesis helps to achieve good sampling of the Fourier
plane. Consequently, imaging planets in this conventional way smears
any structure in longitude because of the rotation of the planet (Jupiter rotates in $\sim$~10 hrs).  In
principle one can merge together snapshots of the same rotational
aspect of the planet from observations taken on different days, if the
structure does not change over time. To image Jupiter's thermal radiation on length-scales
of interest in the atmosphere, one requires a high spatial resolution,
ideally $\sim$~0.5-1\pr. In 5 minutes of time, the rotational smearing
at the disk center is $\sim$~1\pr, and hence integration times should
be of order 1 minute or less. Such images would have such poor
sensitivity that even the zones and belts on Jupiter
may not be recognized.

To overcome these problems, we use an innovative ``facet'' technique to
synthesize together many hours of radio data such that a
longitude-resolved map can be produced.
This technique was developed and applied to 2-cm VLA data of Jupiter by Sault et al. (2004). In short, we effectively approximate the planet by a large number of facets. We apply a phase shift to each visibility datum to account for the motion of a facet across the face of Jupiter. We then apply a linear transformation to the u-v coordinates of the data to correct for changes in the viewing geometry of a facet as it rotates across the disk. Finally we scale up the amplitude (and noise variance) to account for the change in the projected surface area. Each facet is then mapped, where the visibilities are weighted inversely with the noise variance. For best results, we first subtract a limb-darkened disk from the data, so that we only deal with the deviation from this disk (i.e., much smaller amplitudes). In the end we stitch all facets together by regridding them on a common mapping geometry (e.g., Mercator/cylindrical), and feathering them together in regions of overlap.

The method to create longitude-resolved maps works 
best at wavelengths $<$~4 cm, where the influence of Jupiter's synchrotron
radiation is small. At longer wavelengths Jupiter's 
flux density becomes increasingly more dominated by the planet's nonthermal radiation. 
This is in particular noticeable in maps constructed from the low resolution data. The maps in Fig.~\ref{fig1} were created from just the high resolution data.

We used this facet technique in dP16 to create
longitude-resolved maps of Jupiter's thermal emission for each
1-GHz-wide u-v dataset in the Ku, X and C bands, and also by combining
all u-v data in each of the three bands. At the time, for the X and Ku band maps we
used the combination of the high resolution configuration and the short
spacing data up to 12 k$\lambda$ from the low resolution observations; for
the C band data we used only the A configuration data where the
synchrotron radiation is effectively resolved out. In the present paper
we present all the data, and we use only data from the high resolution
configurations. These maps are ``corrected'' for contrast
deficiencies, and the disk-averaged temperature was ``forced'' to be equal to the values in Column 13 of Table 2 (Section~\ref{smeared1}).

The above mapping technique results in a resolution that varies across
Jupiter's disk, and with frequency. An example of a beam pattern is
shown in the Appendix, Figure~\ref{figbeams}.

\section{Radiative Transfer (RT) Model}\label{RT}

We model the data with our radiative transfer code, Radio-BEAR (Radio-BErkeley Atmospheric Radiative transfer); earlier versions were
documented by de Pater et al. (2005; 2014).
Synthetic spectra are obtained by integrating the equation of radiative transfer (RT) through a model atmosphere:

\begin{equation}
B_\nu(T_b) = 2 \times \int_0^1 \int_0^\infty  B_\nu(T) e^{(-\tau_\nu/\mu)} d(\tau_\nu/\mu) d\mu,\label{eq1}
\end{equation}

where the brightness $B_\nu(T_b)$ can be compared to that of the observed disk-averaged brightness temperature, T$_b$. The brightness $B_\nu(T)$ is given by the Planck function and the optical depth $\tau_\nu$ is the integral of the total absorption coefficient from the altitude $z$ to space at frequency $\nu$. The parameter $\mu$ is the cosine of the angle between the line of sight and the local vertical. By integrating over $\mu$, one obtains the disk-averaged brightness temperature, to be compared to the observed brightness temperature, $T_b$.

Before the integration in Equation~\ref{eq1} can be carried out, we calculate the atmospheric structure (composition and TP profile) using an adaptation of the Atreya and Romani (1985) model, with the thermodynamic data as given by Atreya (1986).
We assume the atmosphere to be in thermochemical
equilibrium, and calculate the atmospheric structure after
specification of the temperature, pressure, and composition of one
mole of gas at some deep level in the atmosphere, well below the
condensation level of the deepest cloud layer (typically at a few
10,000 bar). The model then steps up in altitude, in roughly 1 km
steps. At each level, the new temperature is calculated assuming an
adiabatic lapse rate, and the new pressure by using hydrostatic
equilibrium. The partial pressure of each trace gas in the atmosphere
is computed. The criterion for a trace gas to condense
and for a cloud to form from the condensate is that the partial
pressure of the trace gas exceeds its saturation vapor pressure, or
equivalently, that the temperature be below the ``dew point'' of the
trace gas. The temperature then follows a wet adiabat, unless a parameter is set to reduce the latent heat contribution. The condensible gas will follow the saturated vapor pressure curve within and above the cloud layer, unless a relative humidity is specified (e.g., RH$=$10\%).

In Jupiter's
atmosphere we expect an aqueous ammonia solution cloud
(H$_2$O-NH$_3$-H$_2$S), water ice, ammonium hydrosulfide (NH$_4$SH) solid, and
ammonia ice, as indicated in Figure~\ref{figcontrib}. Since the NH$_4$SH cloud forms as a result of a
reaction between one molecule of NH$_3$ with one molecule of  H$_2$S, the test for the NH$_4$SH cloud
formation is that the equilibrium constant of the reaction is
exceeded.  Both NH$_3$ and H$_2$S are reduced in equal molar
quantities until the product of their atmospheric pressures equals the
equilibrium constant. As the trace gases are removed from the
atmosphere by condensation, ``dry'' air (an H$_2$-He mixture) is
entrained into the parcel to ensure the mixing ratios add up to
one. This cycle is repeated until the tropopause temperature is
reached. We choose the temperature and pressure at our base level such
that for every model the temperature is 165 K at the
1 bar level to match the Voyager radio occultation profile (Lindal, 1992). At levels where radiative effects become important ($P \lesssim 0.7$ bar), we replaced the above derived TP profile by one that was determined from mid-infrared ({\it Cassini}/CIRS) observations (Fletcher et al., 2009), and re-calculated the saturated vapor curves to match the new TP profile. We refer to this profile as our nominal TP profile, which follows a dry or wet adiabat at P$\sim$~0.7 bar. Our temperature profile
is shown in Figure~\ref{figcontrib}. 

Equilibrium cloud models typically assume that all cloud material
remains at the altitude level where it condensed (Weidenschilling and
Lewis, 1973). Wong et al. (2015) found that these models are only able
to calculate cloud density formation rates, which then must be
multiplied by an updraft length scale in order to determine the actual
amount of condensed material at each level. This method is good for
calculating fresh cloud densities (before particle growth can lead to
precipitation). For evolved clouds, where eddy diffusion generates new
condensates and precipitation depletes them, the Ackerman and Marley
(2001) approach is more appropriate. Actual cloud densities can vary
by orders of magnitude, depending on the relative efficiencies of
updrafts, diffusion, and precipitation. Given the uncertainties in the
relevant dynamical/microphysical parameters, we performed calculations
neglecting condensed material as an opacity source. Section~\ref{spec} supports this approach.

The gas opacity in Jupiter's atmosphere is primarily determined by
collision-induced absorption due to hydrogen gas (CIA: we include
H$_2$-H$_2$, H$_2$-He, H$_2$-CH$_4$), NH$_3$ and some H$_2$S, while at
longer wavelengths H$_2$O becomes noticeable; at the frequencies used
to observe Jupiter in this paper, ammonia gas is the dominant source
of opacity (e.g., de Pater and Massie, 1985; de Pater et al., 2001;
2005).  Our code has been updated with the new laboratory measurements
for microwave properties of NH$_3$ and H$_2$O vapor, which were
obtained under simulated (high pressure) jovian conditions (Devaraj et
al., 2014; Karpowicz and Steffes, 2011; Bellotti et al., 2016). For CIA we use the absorption
coefficients calculated from revised ab initio models of Orton et
al. (2007), assuming an equilibrium distribution of the hydrogen para
vs ortho states.

Our {\it nominal} model has the following abundances in Jupiter's deep atmosphere: CH$_4$, H$_2$O, and Ar are enhanced by a factor of 4 over the proto-solar values of Asplund et al. (2009)\footnote{The proto-solar values of Asplund et al. (2009) are: C/H$_2 = 5.90 \times 10^{-4}$; N/H$_2= 1.48 \times 10^{-4}$; O/H$_2= 1.07 \times 10^{-3}$; S/H$_2= 2.89 \times 10^{-5}$; Ar/H$_2= 5.51 \times 10^{-6}$.}, and NH$_3$ and H$_2$S are enhanced by a factor of 3.2. This results in abundances (volume mixing ratios) of: CH$_4$: 2.0$\times$10$^{-3}$; NH$_3$: 4.07$\times$10$^{-4}$; H$_2$O: 3.7$\times$10$^{-3}$; H$_2$S: 8.0$\times$10$^{-5}$. The CH$_4$ and H$_2$S values agree well with the abundances derived by Wong et al. (2004a) from the Galileo Probe data, while the NH$_3$ value is close to the lower limit determined by them. We did not include PH$_3$ in our calculations since it has no effect on the spectra discussed in this paper.

In principle, variations in the observed brightness temperature can be caused by
variations in opacity or spatial variations in the physical
temperature. Several authors have looked at the coupling between
differences in temperature between the EZ and NEB, the thermal wind
equation, and observed wind profiles in the cloud
layers. De Pater (1986) considered the difference in latent heat release from the formation of the NH$_4$SH clouds in the EZ (thick cloud layer) and the NEB (thin cloud layer), and determined that this could lead to a temperature difference of $\sim$~3-4 K, matching the value required to drive the zonal winds; however, this would lead to the EZ being warmer than the NEB, which is {\it opposite} to that observed at radio wavelengths. Bolton et al. (2017) determined the difference in physical temperature required to explain the {\it Juno} data (i.e., warm NEB vs cold EZ down to several hundred bar), and calculated the expected zonal wind velocities using the thermal wind equation. They determined that the winds resulting from such a temperature difference would be larger than observed by $\sim$~2 orders of magnitude, and be of the {\it wrong} sign, i.e., corroborating the former conclusion. We therefore assume in this paper that the variations in brightness temperature seen at wavelengths that probe below the cloud layers are caused by variations in opacity rather than physical temperature. As mentioned above, we adopt an adiabat for the TP profile, typically a wet adiabat in a zone and dry in a belt (Fig.~\ref{figcontrib}).

At K band we probe
altitudes right in the middle of the NH$_3$ ice cloud
(Fig.~\ref{figcontrib}), where NH$_3$ gas follows the saturated vapor curve unless it is subsaturated, as has been suggested by several authors based upon microwave data (e.g., Klein and Gulkis, 1978; de Pater et al., 2001; Gibson et al., 2005; Karim et al., 2018), and {\it Cassini}/CIRS observations (Achterberg et al., 2006; Fletcher et al., 2016).

If NH$_3$ gas is fully saturated, the difference in brightness temperature between belts and zones at K band cannot be explained by a difference in physical temperature, since the NH$_3$ abundance is tied to the physical temperature at the altitude of interest. An example is shown in Figure~\ref{figsaturation}, where we show the TP profile with saturated NH$_3$ abundance in panel a, and the resulting spectrum (at the latitude of the EZ) in panel b. As shown, a change in the physical temperature is essentially ``compensated'' by the NH$_3$ abundance, such that the resulting spectra are very similar, both for a TP profile that is warmer (T1, T2), or colder (T3) than the nominal profile. If the NH$_3$ profiles are subsaturated, variations in temperature will effect the spectra; however, the NH$_3$ abundance and TP profile cannot be determined separately from the microwave data alone. Since we have no independent data on the TP profile, we adopt in this paper the nominal TP profile (wet and dry adiabats are quite similar -- see Fig.~\ref{figcontrib}, and derive the spatial variations in the NH$_3$ abundance from the observations. In Section~\ref{CIRS} we evaluate the potential to combine mid-infrared and microwave data to extract both the TP profile and ammonia abundance from the data.

\section{Results}\label{results}

\subsection{Disk-averaged Spectrum}\label{spec}

Figure~\ref{figspec} shows a spectrum of Jupiter's disk-averaged brightness temperature. The VLA data reported in this paper are shown by solid red dots (from Column 13 in Table 2); other data are indicated by other colors/symbols, as summarized in the figure caption. It is interesting to note that our VLA numbers agree very well with the most accurately determined values via other means: $i)$: the {\it WMAP} values ({\it Wilkinson Microwave  Anisotropy Probe satellite}; green points at 1.3, 0.9, 0.74, 0.5, and 0.32 cm, Weiland et al., 2011), $ii)$: Gibson et al.'s (2005) absolutely calibrated datapoint obtained with one of the BIMA antennas (Berkeley-Illinois-Maryland Array; black, at 1.05 cm), $iii)$: the values derived for Jupiter from data obtained with the compact 8-element Sunyaev-Zel'dovich Array (SZA), a subset of CARMA (Combined Array  for Research in Millimeter-wave Astronomy) (cyan datapoints at 0.8--1.1 cm; Karim et al., 2018), and $iv)$: the value derived from data obtained with the radiometer on board the {\it Cassini} spacecraft (Moeckel et al., 2018; solid blue datapoint at 2.2 cm). We therefore think that our adopted calibration uncertainty of 3\% is a conservative value.

Several models are superposed on the spectrum: The blue line is for an
atmosphere with C, N, O, and S abundances according to the most recent
estimates for the proto-solar nebula (Asplund et al., 2009).  The cyan
line is for our nominal model (Section~\ref{RT}), which has a NH$_3$ abundance of 4.1$\times$10$^{-4}$ (3.2 $\times$ solar N) in the deep atmosphere, which decreases at higher altitudes due
to cloud formation (Fig.~\ref{figNH3profiles}a). We assumed a 100\% relative humidity (RH$=$100\%), i.e., a
fully saturated profile.  The magenta line is based on the model of
the cyan line, but where the NH$_3$ abundance has been tweaked to
provide a best fit to the combined CARMA/SZA, {\it WMAP} and BIMA data (from Karim et
al., 2018). This ammonia profile, equal to 5.7$\times$10$^{-4}$ at
P$>$8 bar, decreases to 2.4$\times$10$^{-4}$ above 8 bar, and then to
1.9$\times$10$^{-4}$ above the $\sim$~2.5 bar level due to cloud
formation (Fig.~\ref{figNH3profiles}a).  At altitudes above 0.8 bar ammonia gas is sub-saturated at $\sim$~10\%.
This profile provides an excellent
match to our VLA data.  A comparison between model spectra and data
over the 0.3--3 cm (10--100 GHz) range further suggests that cloud opacity can indeed be
ignored, since such opacities would produce an asymmetry on the two sides of the V-shaped spectrum, with more opacity (resulting in a lower T$_b$) at mm wavelengths compared to cm wavelengths, since cloud particles are most likely micron-- rather than cm--sized.

To give the reader a sense how uncertainties on brightness temperatures would affect the NH$_3$ abundance, we show in dashed magenta lines profiles relative to the magenta line where NH$_3$ has been increased by 20\% at $P \gtrsim$ 1 bar (lower curve), and decreased by 20\% at 1$\lesssim P <$ 8 bar (upper curve).

\subsection{Longitude-smeared Maps}\label{smeared2}

Longitude-smeared maps\footnote{The maps discussed in the remainder of
  the paper all pertain to maps constructed from the high resolution configurations, i.e., A configuration for S and C band,
  B configuration (December 2013 and January 2014) for X and Ku band, C configuration
  for K band. All maps are scaled to the 23 Dec 2013 epoch.} of each full band (Ka, K, Ku, X, C), constructed from
the high resolution configurations after subtraction of a limb-darkened disk as
described in Section~\ref{obs1}, were displayed in Figure~\ref{fig1} and briefly discussed in Section~\ref{smeared1}.
After correction for the total flux and/or brightness contrast in the maps due to missing short spacings (Section~\ref{smeared1}), we reprojected 
each 1-GHz wide map on a longitude/latitude grid. We constructed north-south (NS) scans for each of these maps by median averaging over a longitude range of 60$^\circ$, centered on the central meridian longitude. These scans are shown in Figure~\ref{figscans}. Since a uniform limb-darkened disk had been subtracted from the data, the background level of each scan is centered near 0 K. The y-axis is therefore irrelevant, except to indicate the relative increments in T$_b$. 
In some cases the background level seems to curve upwards at higher latitudes; this could be caused by a slight mismatch in the limb-darkening parameter ($q$ in Table 2). We note that we used the same limb-darkening towards the poles and along the equator, and hence the upturn in T$_b$ suggests that the poles are less limb-darkened than east-west scans along the planet, as shown before from VLA maps (de Pater, 1986) and {\it Cassini} radiometer data (Moeckel et al., 2018). 

At the top we show a visible-light image from the Outer Planet Atmospheres Legacy (OPAL) program on the {\it Hubble Space Telescope (HST)} on 19 Jan. 2015 (Simon et al., 2015).  This helps to connect the radio features to the zone/belt structure in the optical data.
 We have drawn a few vertical lines (dashed), indicated with the numbers 1--7, to highlight some zones and belts. These scans show that the structure of zones and belts is visible at all frequencies between $\sim$~7 and 25 GHz. The variations over time (note that data were taken at three different times spread over a full year) are usually smaller than variations in frequency. The NEB, SEB, and EZ feature prominently even at the lowest frequencies.
Line 1 indicates the minimum brightness temperature in the EZ at $\sim$~4$^\circ$ planetographic latitude, and line 2 is drawn at the planetographic latitude ($\sim$~11$^\circ$) that corresponds to the maximum brightness temperature in the NEB, which is near the north-end of the radio-hot belt. The latter is right at the demarcation line between the brown NEB and a bluish area, indicative of hot spots in the visible. The temperature difference between these two features represents the maximum contrast in each scan. The EZ becomes gradually more and more negative with decreasing frequency compared to the zero-value of the disk, while the NEB peak increases. Several other radio-bright features (e.g., 4, 5, 6, 7, as well as several not indicated by lines) are usually also on latitudes separating differently colored regions in the visible, while line 3 falls at a narrow zone.

Figure~\ref{figdeltaTb} shows the peak-to-peak difference in brightness temperature between the NEB and EZ. We see a steady increase in $\Delta$T$_b$ with decreasing frequency between 20 and 3 GHz; however, the values below 5 GHz become more unreliable due to a combination of missing short spacings (although this was partially corrected, Section~\ref{smeared1}) and a relative increase in contamination by synchrotron radiation. 
For comparison, we superpose the $\Delta$T$_b$ from the Microwave Radiometer (MWR) PJ1 {\it Juno} data (red points; Li et al., 2017), taken on 27 August 2016 along one orbital pass. These values are for nadir viewing (i.e., will be slightly higher), and not averaged over longitude, in contrast to the VLA values. The comparison is exceptionally good. Not surprisingly, discrepancies are largest at 10 and 22 GHz, a frequency range where longitudinal variations are largest (dP16; Section~\ref{resolved2}). The exceptionally large {\it Juno} value at 0.6 GHz might in part be caused by synchrotron radiation leaking into the sidelobes of MWR's beam pattern; at this frequency Jupiter's synchrotron radiation is much larger than its thermal emission.

The uncertainties on the VLA data points were derived from the standard deviation over 1$^\circ$ in latitude; the C and S band errorbars were increased by an extra 10\% to account for the large uncertainties in the factors ``adopted'' to correct for the brightness contrast in the maps (Section~\ref{smeared1}).

The NEB appears in Figure~\ref{figscans} as a bright feature, extending northwards of dashed line 2. At high frequencies (22--25 GHz), probing aproximately the NH$_3$ cloud deck ($\sim$~0.4--1 bar), the NEB is quite broad; at $\sim$~13--17 GHz, probing $\sim$~0.5--3 bar, the NEB shows a pronounced peak at $\sim$~11$^\circ$ latitude, and at lower frequencies probing deeper layers in the atmosphere, the NEB broadens considerably towards higher latitudes (see line 3 at $\sim$~22$^\circ$), and the peak brightness temperature of the NEB may even shift from $\sim$~11$^\circ$ to 20$^\circ$ shortwards of 4 GHz (below the 2--3 bar level). This agrees with findings obtained by {\it Juno}/MWR (Bolton et al., 2017; Li et al., 2017).

Another pronounced belt in the K band (18--25 GHz) is seen at line 4 in Figure~\ref{figscans}; this belt, though visible from 25 down to 16 GHz, appears to disappear at lower frequencies, i.e., the brightness temperature at lower frequencies is not any higher than surrounding latitudes. In contrast, line 3 coinciding with a zone at 25--18 GHz, looks more like a belt at lower frequencies ($\sim$~5--9 GHz). Since K band data were taken at a different time, we cannot rule out time variability here, however. But if true, the ``switch'' from a zone-like band in the upper atmosphere to a belt-like band deeper down must be caused by dynamics, such as a reversal of circulation patterns, leading either to a decrease in the NH$_3$ abundance, and/or an increase in the physical temperature in the deeper atmosphere. Such reversals in circulation patterns have been suggested before for Jupiter (Ingersoll et al., 2000; Showman and de Pater, 2005; Marcus et al., 2019), Saturn (Fletcher et al., 2011), and Neptune (Tollefson et al., 2018). Other pronounced zones and belts are visible in the southern hemisphere (e.g., such as indicated by lines 5--7), several of which are quite prominent both at high and at low frequencies (such as along lines 5 and 6). Although the data were taken at different times, the zone-belt structure appears to vary more with frequency than over time. We do need to bear in mind, though, that this analysis was performed on longitude-smeared maps, and as shown in Section~\ref{resolved2}, there is much longitudinal structure which will affect the structure of the scans.

\subsection{Longitude-resolved Maps}\label{resolved2}

Longitude-resolved maps at the Ka, K, Ku, X, C and S bands are shown in Figure~\ref{figresolved}. A best-fit limb-darkened disk was subtracted from each of the maps (Section~\ref{obs1}), and hence bright areas indicate a higher brightness temperature, assumed to be caused by a lower NH$_3$ abundance, and dark areas indicate a lower brightness temperature, caused by a higher opacity in the atmosphere. A wealth of structure is visible in each map, as briefly summarized below (and in dP16). It is easy to distinguish this structure from ripples and large-scale dark and light areas that are instrumental artefacts (such as indicated by the green arrows).

{\it Ku (12-18 GHz) band}: In the Ku band (probing roughly 0.5--3 bar) we see a lot of structure over the entire globe. The radio-hot belt at 8.5--11$^{\circ}$ latitude features prominently,  with hot spots (some of which are indicated by yellow arrows) interspersed with small well-defined dark regions, probably small plumes of ammonia gas. Just to the south are larger dark somewhat oval-shaped regions; these are the plumes of ammonia gas discussed before (dP16), which are most prominent in C band (red arrows).
The Great Red Spot (GRS) and Oval BA are well-defined structures surrounded by a bright ring, and they have a small brighter dot (higher T$_b$) at their center. Turbulent wakes are visible on the west side of both storm systems. 
Small scale structures are spread all over the globe. For example, numerous small and larger vortices are visible (such as those indicated by cyan arrows), typically characterized by a darker center surrounded by a brighter ring, similar to the small vortices seen at a wavelength of 5-$\mu$m (de Pater et al., 2010; 2011). Many of the same structures are seen in visible wavelength data (see comparisons in dP16). More features can be distinguished in the December than in the January data, due to a higher data quality and higher spatial resolution at the time.

{\it X (8-12 GHz) band}: At X band we see very similar structures as seen in Ku band, though slightly subdued, caused in part by a lower spatial resolution (Table 3) and perhaps by the fact that slightly deeper layers in the atmosphere are probed, down to $\sim$~8 bar (Fig.~\ref{figcontrib}). 

{\it Ka (29--37 GHz) band}: As mentioned above, the spatial resolution in the Ka band is much lower than that at other bands (Table 3), yet the hot spots, plumes, and the GRS are clearly visible. This is not surprising since the weigthing functions at Ka and Ku bands probe the same altitudes (Fig.~\ref{figcontrib}).

{\it K (18-26 GHz) band}: Numerous features are visible at K band, a wavelength region where we probe Jupiter's NH$_3$ ice cloud (Fig.~\ref{figcontrib}), and that is especially sensitive to the NH$_3$ humidity in and above this cloud. The GRS is clearly visible, with a very bright ring on the south side and turbulent wake to the west. A bright ring on the south side was also seen at mid-infrared wavelengths (Fletcher et al., 2010), and interpreted to be caused by a high physical temperature at 500--800 mbar. We will discuss this feature further in Section~\ref{resolved3}. Oval BA is a dark oval south-east of the GRS; the wake is hardly visible. 
Clearly visible are the radio-hot belt with hot spots and other fine-scale structures (bright/dark patterns), and plumes of ammonia gas just to the south of it.

\begin{figure*}
\includegraphics[scale=0.60]{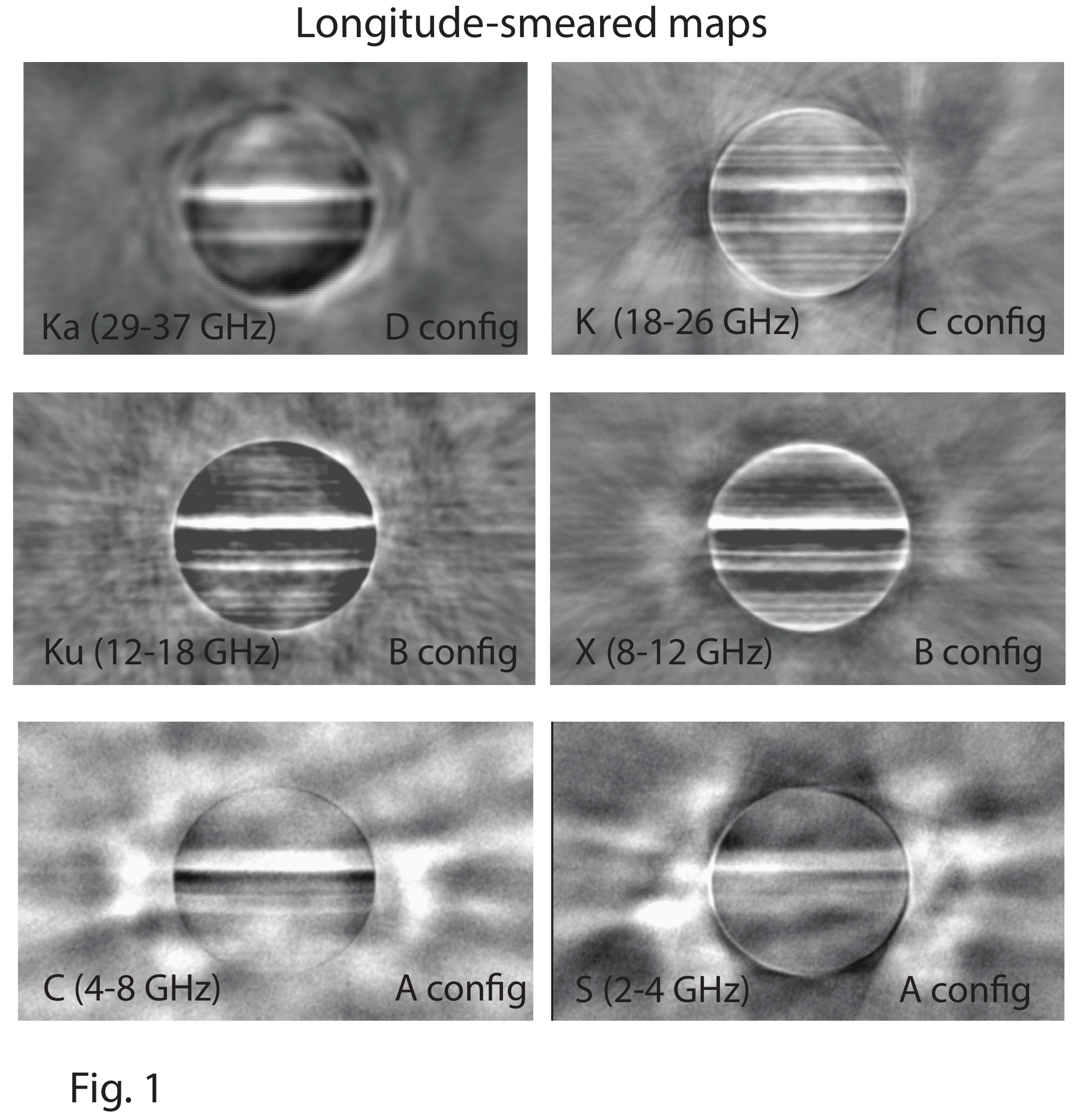}
\caption{Longitude-smeared maps of Jupiter's thermal emission in all 6 observing bands. The maps were produced using data from only the extended array configurations; for the B configuration we used data from only December 2013. The spatial resolution in all maps is 0.8'', except for the Ka map, where it is 2.5''. Further details are provided in Tables 1 and 2. }
\label{fig1}
\end{figure*}

\begin{figure*}
\includegraphics[scale=0.37]{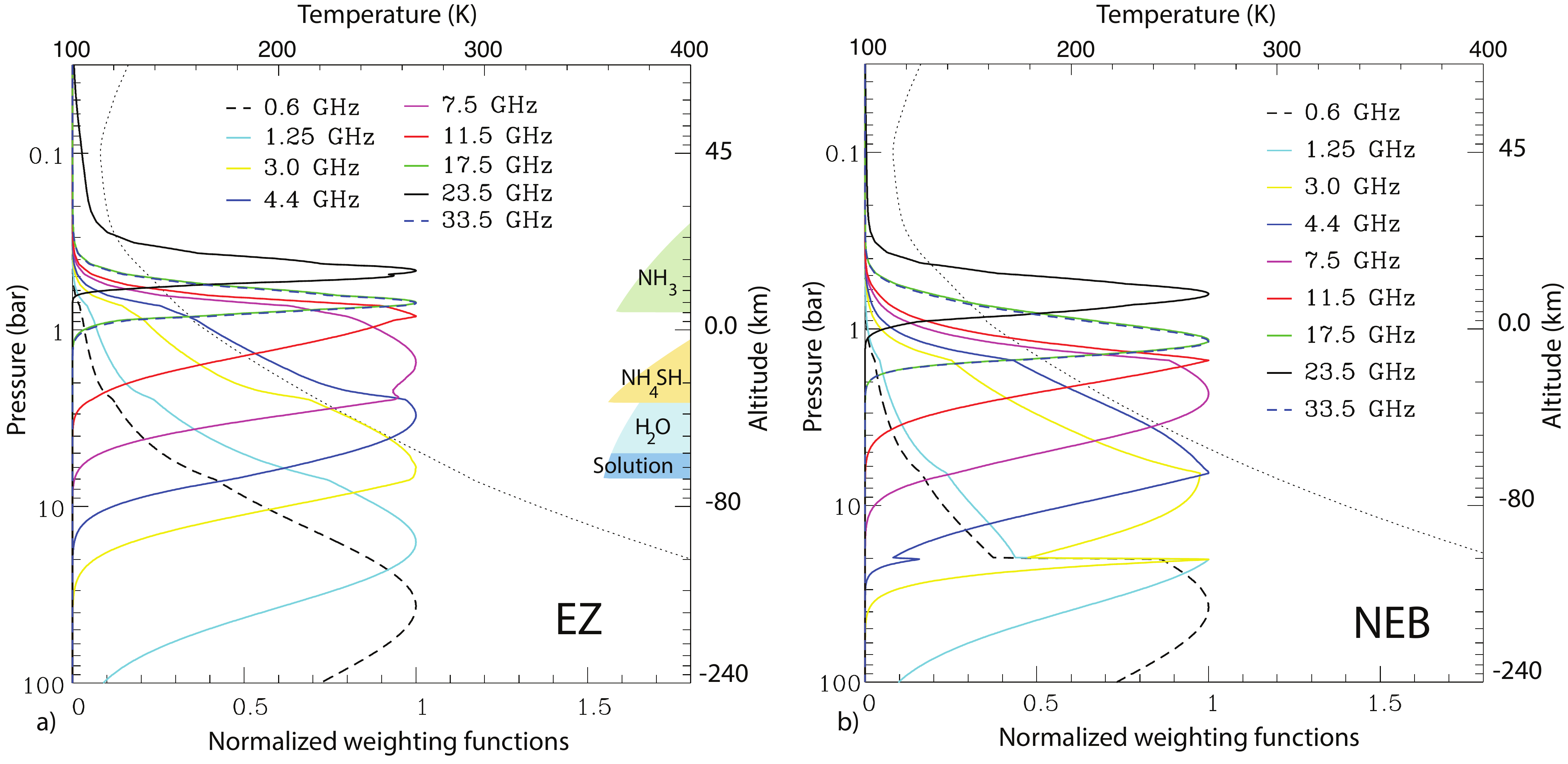}
\caption{a) The temperature profile (dotted line; wet adiabat) and weighting functions for our nominal atmosphere in thermochemical equilibrium, under nadir viewing conditions. In our nominal atmosphere the abundances of CH$_4$, H$_2$O, and Ar in the deep atmosphere are enhanced by a factor of 4 over the solar values, and NH$_3$ and H$_2$S are enhanced by a factor of 3.2 each. On the right-hand side we indicate the various cloud layers expected to form in thermochemical equilibrium. b) The temperature profile (dotted line; dry adiabat) and weighting functions for the NEB, where NH$_3$ is decreased in the troposphere, as indicated in Fig.~\ref{figNH3profiles}, under nadir viewing conditions. Closer to the limb the profiles will shift upwards. Note that the weighting functions at 33.5 and 17.5 GHz essentially coincide. Although weighting functions at frequencies as low as 0.6 GHz are shown, we note that the thermal emission as observed from Earth will be increasingly compromised with decreasing frequency below 3 GHz.
}
\label{figcontrib}
\end{figure*}

\begin{figure*}
\includegraphics[scale=0.65]{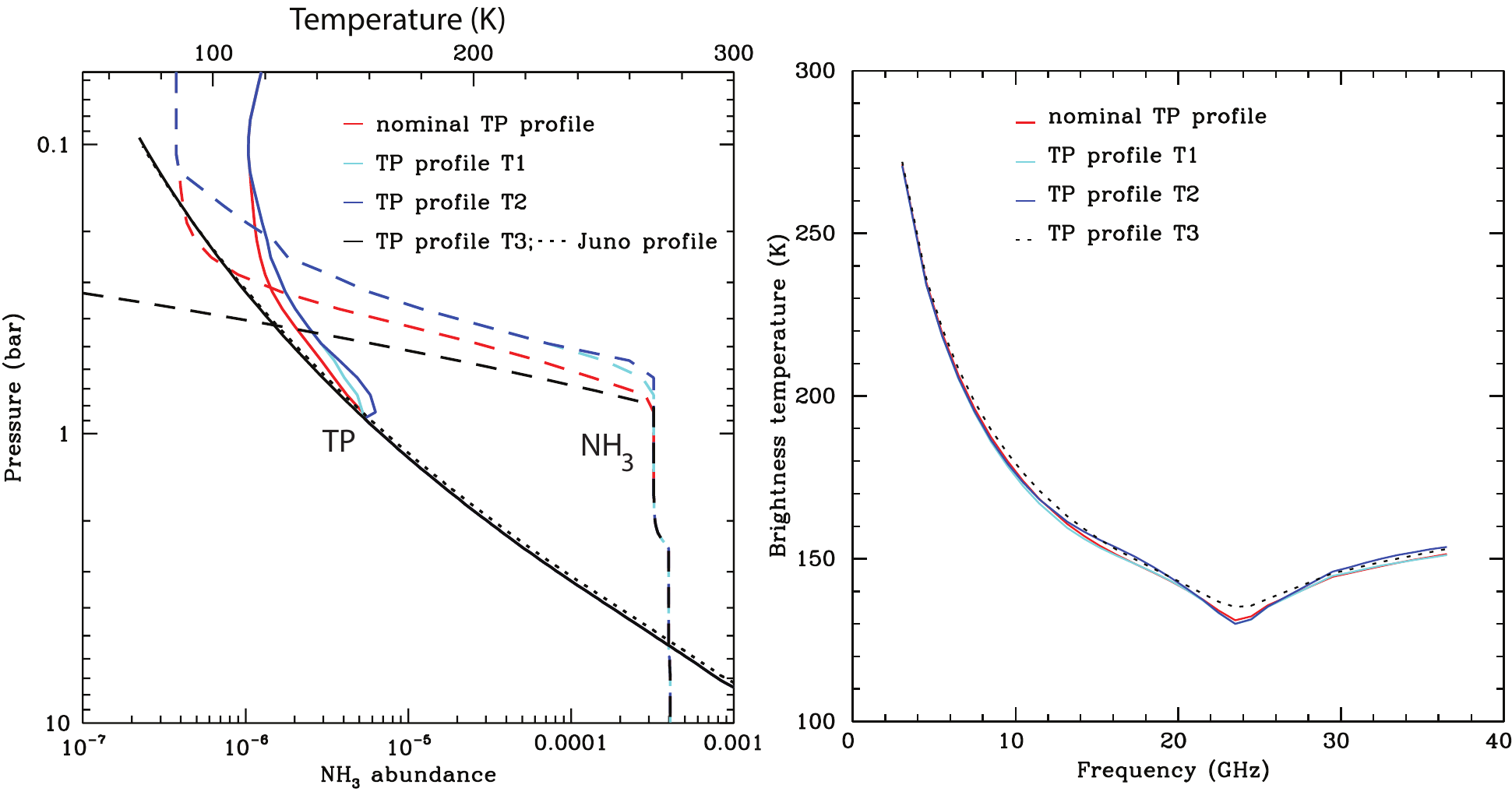}
\caption{Model calculations for different TP profiles and saturated vapor pressure curves for NH$_3$ (panel a) and resulting spectra (panel b). These spectra were calculated for nominal (100\% saturated) profiles at a planetographic latitude of 4$^\circ$. Within the NH$_3$ cloud layer (where temperature and saturated NH$_3$ vapor pressure are correlated), changing the temperature profile has only a slight effect on the microwave spectrum. The dotted TP profile is the profile used by the {\it Juno} team, which is essentially a simple adiabat up to the tropopause (Li. et al., 2017). }
\label{figsaturation}
\end{figure*}

\begin{figure*}
\includegraphics[scale=0.60]{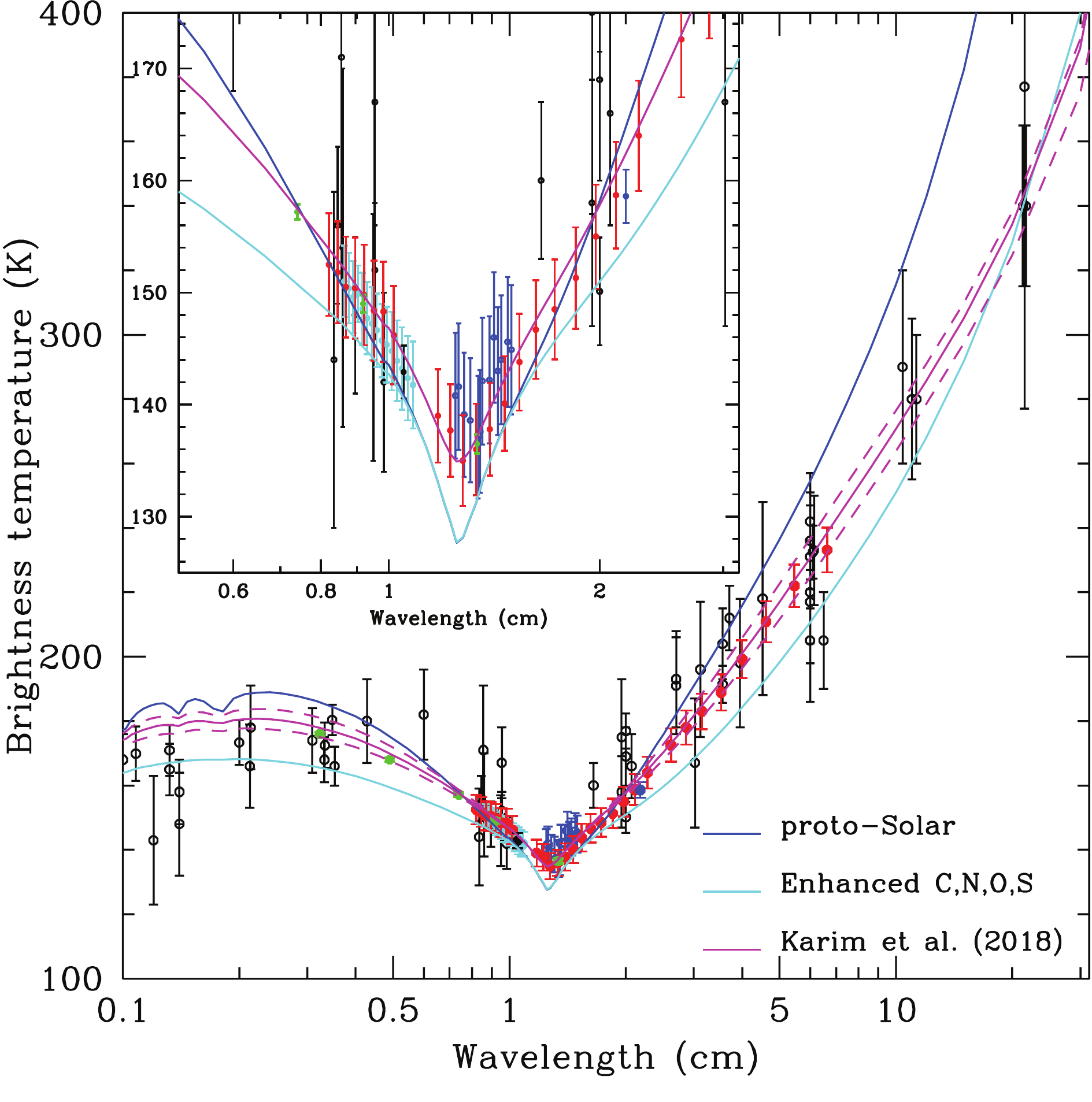}
\caption{Jupiter's disk-averaged brightness temperature as a function of
wavelength. The filled red dots with errorbars are the values reported
in this paper (Table 2). 
The open black symbols are older datapoints, as compiled by de Pater et al. (2005). The open blue circles near 1.3 cm (Klein et al., 1978), the filled green dots (Weiland et al., 2011), and the black datapoint at 1 cm (Gibson et al., 2005) were corrected according to the procedures outlined in Gibson et al. (2005). The cyan data points at $\sim$~0.8--1 cm are from Karim et al. (2018), and the blue datapoint at 2.2 cm is from Moeckel et al., (2018). The various curves are model calculations: the blue curve is for a jovian atmosphere with solar abundances for all gases; the cyan curve is our nominal model where NH$_3$ and H$_2$S are enhanced by a factor of 3.2 compared to the solar values, and H$_2$O and CH$_4$ by a factor of  4; the magenta curve is from Karim et al. (2018). The dashed magenta spectra show the effect of an increase/decrease in NH$_3$ on the latter curve. For the lower dashed line we increased the NH$_3$ abundance by 20\% (multiplied by 1.20) at $P \gtrsim 1$ bar; for the upper dashed line we decreased NH$_3$ by 20\% (divided by 1.20) at $1 \lesssim P \lesssim 8$ bar. The inset shows a zoomed-in view of the center part of the spectrum (without the dashed magenta curves). 
}
\label{figspec}
\end{figure*}

{\it C (4-8 GHz) and S (2-4 GHz) bands}: At longer wavelengths the picture starts to change. At C band, probing down to over 10 bar, the wave-train of NH$_3$ plumes is striking (indicated by red arrows), as discussed before by dP16, while the radio-hot belt at 8.5--11$^{\circ}$ latitude is quite prominent as well. Small-scale structures in this belt are visible in particular in the east, north of the GRS. At $\sim$~20-21$^\circ$N another faint bright band is visible, which caused the broadening of the NEB in latitude in the longitude-smeared images (Figs.~\ref{fig1},\ref{figscans}). The GRS is still quite visible, with a darker center surrounded by a brighter ring and a wake to the west. Oval BA itself, in contrast to its wake, nor smaller vortices can be distinguished. This is not caused by a lower angular resolution; the full angular resolution is similar to that in the Ku band (dP16). The maps presented here have been slightly degraded in resolution compared to those presented in dP16 to increase the signal-to-noise, and is only slightly lower than that in X band (Table 3). It does mean that the small-scale dynamics producing these vortices appears to be mostly confined to the ``weather'' layer, the upper 2--3 bar of the atmosphere where the NH$_4$SH and NH$_3$ clouds form (dP16). We say ``perhaps'', since we use spatial variations in the NH$_3$ abundance as tracers of dynamics, assuming that in parcels of rising air, NH$_3$ may be relatively enriched compared to neighboring regions, while descending flows that originate at or above the NH$_3$ cloud base carry depleted ammonia with respect to the deep abundance.
Zones and belts are still visible in C band, though, and hence extend deeper in the atmosphere. One may also notice that the bright bands, i.e., the belts, are often not coherent structures in any of these maps, in particular in the southern hemisphere, where they often show up as series of segments, like a wave pattern. 

The faint large-scale broad dark and bright pattern visible from the north-west to the south and up to the north east is caused by Jupiter's synchrotron radiation. As shown, though, at C band this emission is not too much of a problem. It is much more pronounced at S band. We were able to largely suppress the synchrotron emission at S band by modeling the emission and subtracting it from the map (the original map is shown in the inset). The main take-aways from the S band map are that the plumes are very prominent; the NEB is clearly visible and extended in latitude, while the SEB is much fainter (as in the longitude-smeared maps, Figs.~\ref{fig1},\ref{figscans});  the GRS at W. longitude $\sim$~30$^\circ$ can just be recognized, but is much fainter than in the C band map.

\section{Analysis and Discussion}\label{discussion}

\subsection{Longitude-smeared Maps}\label{smeared3}

To facilitate analysis of the latitudinal structure, we used the north-south scans in Figure~\ref{figscans}  to create a map of the brightness temperature as a function of frequency (y-axis) and latitude (x-axis); this map is displayed in Figure~\ref{figSCXUK}, top panel. Although this map spans the frequency range from 3 up to 37 GHz, we note that the Ka band data (29--37 GHz) have a much lower spatial resolution. This map is used to determine the altitude profile of the NH$_3$ abundance at each latitude via radiative transfer (RT) calculations with our Radio-BEAR code (Section~\ref{RT}).
Our final model fits are shown in the bottom panel of Fig.~\ref{figSCXUK}. We note that our analysis is anchored on the disk-averaged T$_b$ in column 13 of Table 2. As discussed in Section~\ref{spec}, the uncertainties reported in Table 2 are conservative. Figure~\ref{figspec} shows that if the spectrum would deviate from the listed values, the derived NH$_3$ profiles might change perhaps by $\sim$~10\%, likely no more than 20\%. 
If the disk-averaged spectrum would be slightly off, spectra at individual latitudes and of specific features would be off by the same amount, i.e., the NH$_3$ profiles will not change relative to each other but all go up or down by the same amount. We therefore do not list uncertainties in the derived NH$_3$ profiles.

We generated over a 100 different model maps. In each model we used either a dry or wet adiabat (this did not noticeably change the spectra; see Fig.~\ref{figcontrib} for the difference between dry and wet profiles), and a different NH$_3$ profile. These profiles were chosen such as to cover a large grid of potential NH$_3$ abundance vs pressure profiles. 
All models have in common that the NH$_3$ abundance never increases with altitude; they either stay constant, or decrease. Although the profiles derived from the {\it Juno} data showed a minimum in the NH$_3$ abundance at $P \sim$ 5--7 bar, with high values at the bottom of the NH$_3$-ice cloud, we initially considered more ``conservative'' or physically plausible abundance profiles. In Section~\ref{juno} we compare our results with a spectrum derived from a {\it Juno}-derived NH$_3$ profile (Fig.~\ref{figjunoprofile}). 
 
For the final profiles we adopted a value of 4.1$\times$10$^{-4}$ (3.2 $\times$ solar N) in the deep atmosphere at all latitudes (as in our nominal model), usually at pressures $P>$ 8-10 bar, in some cases only at $P>$ 20 bar. This value was based upon best fits to C band data. We ran both chi-square fits of the models to the data at each latitude, using frequencies between 4 and 25 GHz (i.e., C, X, Ku and K bands only), and examined the fits by eye to make sure the chosen models did indeed fit the data. We did not use the 29--37 GHz data in the fits as these have a spatial resolution that is $\sim$~4 times lower than that in the other data. The 3 GHz (S band) values were not used due to the larger uncertainties in this band, attributed mainly to contamination by synchrotron radiation. We show a selection of fits in Figure~\ref{figmodelfits}. In each panel we show in addition to our best fit, also the models that best fit the EZ (our nominal model with a 50\% relative humidity at $P < 0.5$ bar; see Fig.~\ref{figNH3profiles}) and the NEB, which together represent two extremes; the latter curves were plotted in each panel at the emission angle appropriate for the latitude in that panel. 

The NH$_3$ altitude profiles for several of the fits are shown in Figure~\ref{figNH3profiles}a, and Figure~\ref{figNH3} shows a map of the  NH$_3$ abundance as a function of altitude (pressure) at all latitudes. As mentioned above, we adopted an abundance of $\sim$~4 $\times$10$^{-4}$ at $P>$ 8 bar, but at a few latitudes (in several belts) we had to extend the low NH$_3$ abundance down to deeper levels; in the NEB we had to extend it down to the 20 bar level in order to fit the data. At the top of the NH$_3$ abundance panel we show the approximate location of the various cloud layers. The NH$_3$ abundance clearly drops at all latitudes at the 0.7--0.8 bar level, the base level of the NH$_3$-ice cloud, unless the abundance was already less than the saturated vapor curve. Moreover at all latitudes the abundance profile within/above the NH$_3$-ice layer is subsaturated, likely caused by photolysis at the higher altitudes ($P \lesssim 0.3$ bar), and dynamics deeper down. The formation of the NH$_4$SH layer near the 2.2 bar level is also prominent, in particular at mid-latitudes. This cloud layer is expected to form based upon thermochemical arguments (Weidenschilling and Lewis, 1973). Although no direct evidence of this cloud layer exists, the absence of H$_2$S in Jupiter's spectrum at 2.7 $\mu$m (Larson et al. 1984) shows that this gas has a major loss process somewhere below the NH$_3$ cloud layer, which most likely is through the formation of NH$_4$SH. Although sharp, definitive infrared spectral signatures of solid NH$_4$SH have not been found, Sromovsky and Fry (2010a,b) detected broad absorption from solid NH$_4$SH in Jupiter's 3-$\mu$m spectrum as observed by {\it ISO} and {\it Cassini}/VIMS. Most recently, Bjoraker et al. (2018) found evidence for this cloud in the GRS, based upon a joint analysis of groundbased Keck/IRTF 5-$\mu$m spectra at high spectral resolution and {\it Cassini}/CIRS data. Amongst our many model maps, we had some that did include the formation of this NH$_4$SH layer, and many that did not. The fact that models including this cloud, i.e., models that preferred an extra loss of NH$_3$ at this altitude, gave a best fit to observed spectra at many latitudes provides further evidence for the presence of this cloud.

At latitudes $\theta \gtrsim |50|^\circ$, the ammonia abundance at $P>$ 0.8 bar is smaller than at the lower latitudes, in agreement with previous work (de Pater, 1986; Moeckel et al., 2018).  The middle panel shows a copy of Figure~\ref{figSCXUK}, with a north-south scan from Figure~\ref{figscans} superposed. The right panel shows a slice through a visible-light map (from Simon et al., 2015) to emphasize the zones and belts in the atmosphere.
To help interpret the abundance values quantitatively, several slices through the NH$_3$ abundance map are plotted in Figure~\ref{figNH3slice}. The NH$_3$ abundance at 4$^{\circ}$N in the EZ is much higher than anywhere else on the planet, while the NEB at $\sim$~11$^\circ$ shows the lowest abundance. At the 0.5 bar level the humidity varies from RH $\sim$ 50\% in the EZ to  $\sim$~1\% in the NEB. Note, though, that the RH values would change if the TP profile would be different, as discussed in Section~\ref{RT} --  i.e., if the temperature over the EZ is a few K higher, NH$_3$ might follow the saturated vapor curve.

\subsubsection{Comparison with {\it Juno} data}\label{juno}

It is interesting to compare our VLA data to {\it Juno}'s Perijove 1 (PJ1) data (Bolton et al., 2017; Li et al., 2017). Since the {\it Juno} data are all at nadir viewing, we cannot compare our data directly, except near the equator. Instead, Figure~\ref{figmodelsjuno} shows the models that best fit our VLA data (Figs.~\ref{figmodelfits},\ref{figNH3},\ref{figNH3slice}) after conversion to nadir viewing, superposed on the {\it Juno} data. The full PJ1 spectrum is shown, and the frequency range overlapping with the VLA data is enlarged in the inset.  As in Figure~\ref{figmodelfits}, in the inset we show both the best fit VLA profile at that latitude, as well as the best fits to the EZ (cyan) and NEB (blue), all converted to nadir viewing. As shown, our best-fit VLA models match the PJ1 data extremely well, even at frequencies below 4 GHz, the lowest frequency used in fitting VLA data. Only at 0.6 GHz ({\it Juno} Channel 1) do we see a discrepancy, in that our modeled brightness temperatures  are somewhat low compared to the data. However, we note that any leakage of synchrotron radiation through sidelobes of the beam pattern would raise the brightness temperature in channel 1 the most, and the values as published may be on the high side. The match between PJ1 data and the models between 1 and 24 GHz is quite stunning, in particular when realizing that the models shown {\it have not been fitted to PJ1 data}, but to longitude-averaged brightness temperatures from VLA data taken 2 years earlier. As shown in the present paper (Fig.~\ref{figresolved}) and before (dP16; Bolton et al., 2017), longitudinal variations in the brightness temperature are substantial, in particular at frequencies 8--25 GHz. This might explain the tiny discrepancies at these frequencies seen at, e.g., 18$^\circ$ S an 18$^\circ$ N.

This excellent match does raise a question about the uniqueness of the ammonia distribution that the {\it Juno} team obtained by inverting the PJ1 data (Li et al., 2017; Bolton et al., 2017). We refer here in particular to their low NH$_3$ abundance near the 5--7 bar level and their relatively high abundance just below the NH$_3$ cloud layer. We do note that, in particular in K band (and MWR channel 6), use of an accurate temperature-pressure profile is important, which is the reason we use the TP profile as derived from mid-infrared data (Fletcher et al., 2009) at altitudes where the atmosphere is in radiative-convective equilibrium ($P \lesssim 0.7$ bar). Janssen et al. (2017) and Li et al. (2017) adopted a simple adiabat throughout the atmosphere, shown by the dotted line in Figure~\ref{figsaturation}, which results in a tropopause temperature that is several tens of K below the observed value. As shown in Section~\ref{RT}, for a fully saturated NH$_3$ profile the precise TP profile is not very important for calculations of the brightness temperature; however, as mentioned in Section~\ref{RT}, microwave and {\it Cassini}/CIRS data have shown that NH$_3$ gas is globally subsaturated within and above the NH$_3$ cloud layer, and hence knowledge of the precise TP profile is needed to derive the NH$_3$ relative humidity at these altitudes. 

To investigate a possible increase in altitude in the NH$_3$ abundance at $P \lesssim$ 5-6 bar, as published by the {\it Juno} team (Bolton et al., 2017; Li et al., 2017), Figure~\ref{figjunoprofile}  shows a comparison of spectra based upon the NH$_3$ profile derived by the {\it Juno} team from the {\it Juno} data, with a spectrum based upon the NH$_3$ profile derived by us from the VLA data. These NH$_3$ profiles, at a latitude of 28$^\circ$, are shown in panel a), together with our best fit profiles for the EZ and NEB, which provide brightness temperature spectra at the two extreme ends (i.e., all our spectra lie in between these two extremes), as also used in our other figures. Panels b)--d) show the spectra resulting from these NH$_3$ profiles, at the appropriate latitude (28$^\circ$) and viewing geometry (i.e., nadir for the {\it Juno} data). Panels e) and f) show the difference in brightness temperature between the VLA \& {\it Juno} data and the models based on the NH$_3$ profiles as derived from the VLA (magenta curves) and {\it Juno} (blue curves) spectra, $\Delta T_b = T_b({\rm data}) - T_b({\rm model})$. Since we look at differences rather than absolute values, we can plot the {\it Juno} (nadir viewing) data derived from panels b) and d) on the same plots as the VLA data derived from panel c). As shown, $\Delta T_b$ is similar in value for the {\it Juno} and VLA data, and significantly larger when compared to the {\it Juno}-derived model than for the VLA-derived model. We therefore conclude that we do not see evidence for a minimum in the NH$_3$ abundance at $P \sim$ 5--6 bar, with an increase in NH$_3$ abundance at higher altitudes. We note that the {\it Juno} data at frequencies below 3 GHz provide a better match to the spectrum using the {\it Juno}-derived NH$_3$ profile because of its lower NH$_3$ abundance at $P \gtrsim$ 10 bar (Fig.~\ref{figjunoprofile}a). (Note that we only used data between 4 and 25 GHz in our fits; in a future paper we plan to model VLA and {\it Juno} data simultaneously, using datasets taken at the same time).

\subsubsection{Combining VLA and {\it Cassini}/CIRS data}\label{CIRS}

Fletcher et al. (2016) analyzed {\it Cassini}/CIRS mid-infrared data, which are sensitive to the altitude range between approximately 1 and 700 mbar. They derived the TP profile in combination with the aerosol opacity, and the ammonia, phosphine, ethane, and acetylene mixing ratios at each latitude (in 2$^{\circ}$ increments), averaged over longitude. There is a degeneracy, however, in deriving so many parameters from one set of observations. If we could combine microwave and mid-infrared data, we can break the degeneracy between temperature and NH$_3$ abundance.  
To evaluate the potential of such a combined mid-infrared and microwave analysis, we constructed spectra based upon our best fit models, where we replaced the TP profile and NH$_3$ abundance in the upper atmosphere (at $P \lesssim 0.6$ bar) with the CIRS profiles. Results are shown in Figure~\ref{figmodelsCIRS}. Panel a) shows the TP and NH$_3$ profiles in the EZ and NEB as derived from the CIRS data at $P \lesssim$ 0.6 bar (based on Fletcher et al., 2016), in comparison with our nominal profiles. These profiles show large variations from latitude to latitude, as shown in panel b), where the latitudinal variation in CIRS-derived temperature and NH$_3$ abundance is shown at a pressure level of 500 mbar. The resulting spectra are shown for various latitudes in panels c--h). Since we used the best fit VLA models at lower altitudes, the cyan (CIRS-based at $P \lesssim$ 0.6 bar) and red (VLA-based) lines coincide at freqencies below 18 GHz, and above 30 GHz. At frequencies where we probe the NH$_3$-ice cloud and above ($\sim$~18--26 GHz), slight differences can be seen at some latitudes, e.g., in the NEB (12$^{\circ}$ N) and at 16$^{\circ}$ S; but at many other latitudes the profiles are well-matched. The CIRS data were taken in December 2000, 14 years before the VLA data were obtained. As shown by Fletcher et al. (2016), the TP and NH$_3$ profiles may change considerably over time. Although the IRTF/TEXES data shown by them were taken in the same year as the VLA data, we opted not to use those since the results are not as reliable as those from the CIRS data (due to calibration uncertainties). The take-away point from the above exercise is that we can break the degeneracy between ammonia and temperature in the upper troposphere where the contribution functions overlap, which also provides more stringent constraints for the extrapolations to greater depth. To date, there have been no attempts to make the analyses between mid-infrared and microwave data consistent with one another, which ideally would require simultaneously obtained datasets.

\subsection{Longitude-resolved Maps}\label{resolved3}

\begin{figure*}
\includegraphics[scale=0.75]{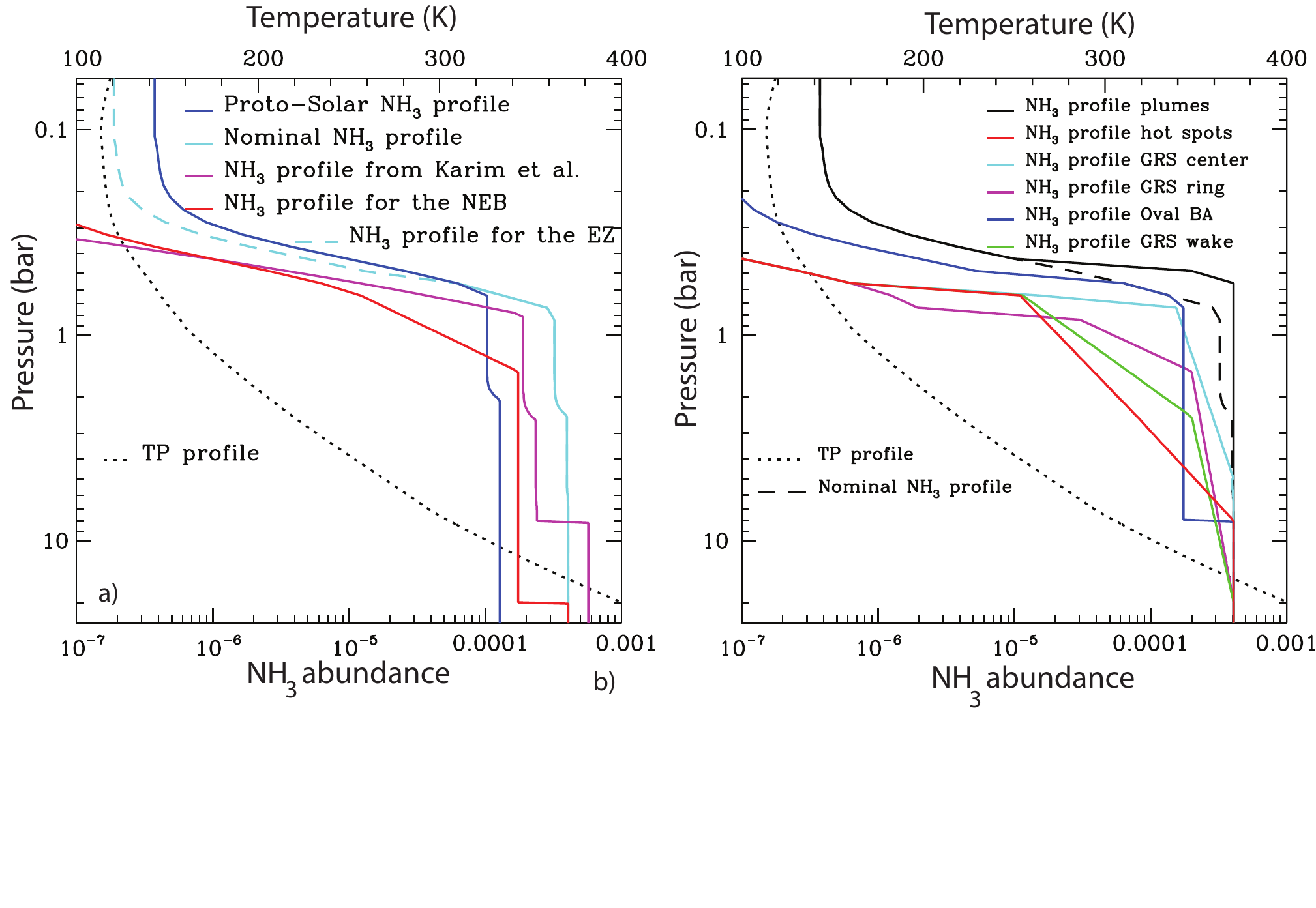}
\caption{a) NH$_3$ altitude profiles for our nominal (and EZ -- dashed) model (cyan), a solar composition atmosphere (blue), a best fit to the CARMA/SZA, {\it WMAP} and BIMA data combined (magenta; from Karim et al., 2018), and a best fit to the NEB. The nominal and solar composition models are fully saturated; the EZ model has a relative humidity RH $=$ 50\% at P$<$ 0.5 bar; the Karim et al. curve is equal to 2.6$\times$10$^{-8}$ at the tropopause and above, while the the NEB profile has a RH$=$1\% (i.e., is equal to 3.75$\times$10$^{-8}$ at the tropopause and higher altitudes).
b) NH$_3$ altitude profiles for several features in the longitude-resolved maps. The dashed line is our nominal model, which is fully saturated, and hence the plume models show supersaturation. The modelfits to the data are shown in Figs. 10 and 16. 
}
\label{figNH3profiles}
\end{figure*}

\begin{figure*}
\includegraphics[scale=0.6]{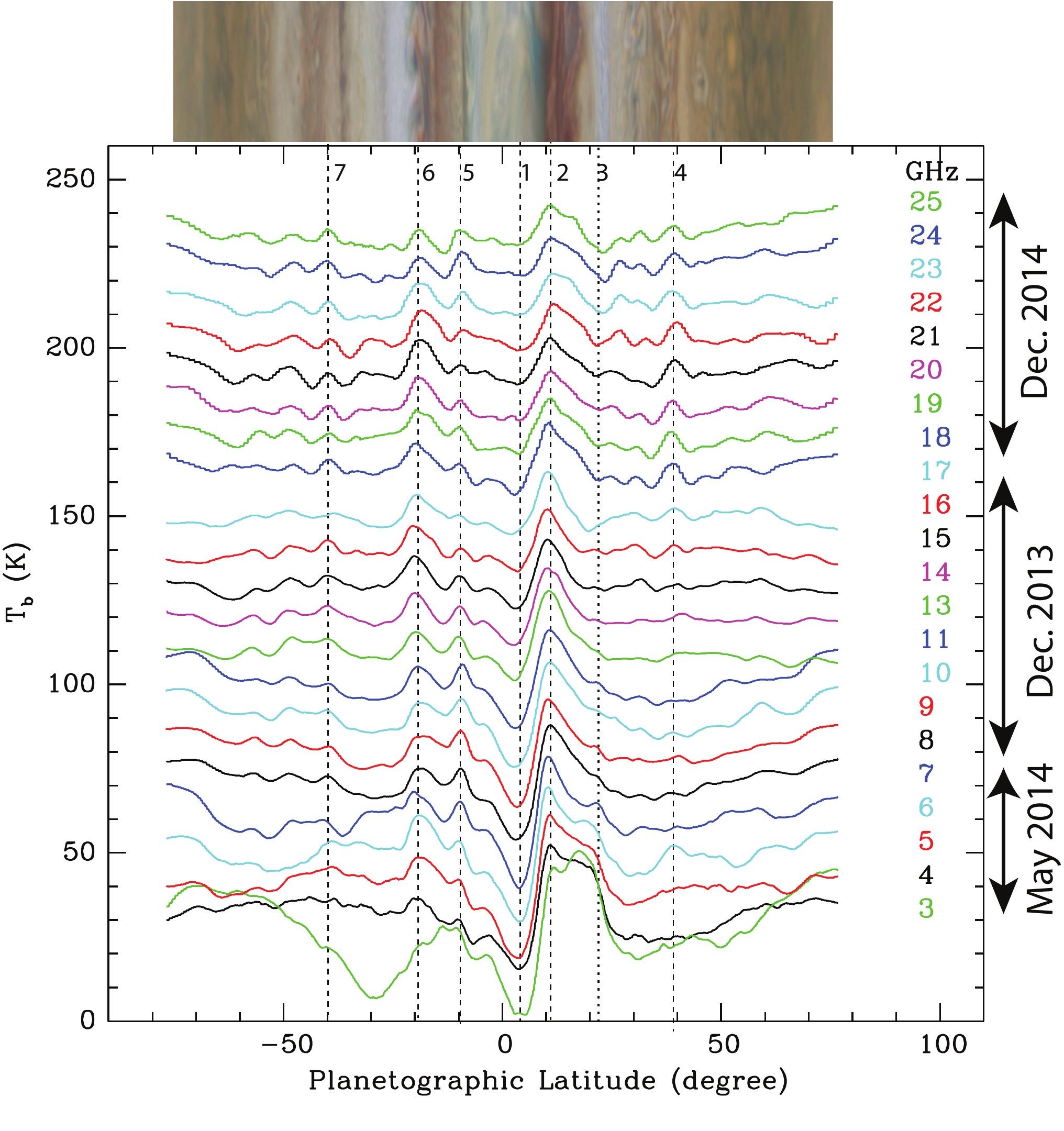}
\caption{North-south scans through longitude-smeared maps of each 1-GHz wide chunk of data, after subtraction of a uniform limb-darkened disk (such as those displayed in Fig.~\ref{fig1}). The scans were created by median averaging over 60$^\circ$ of longitude, centered on the central meridian of each map, after reprojection on a longitude/latitude grid. Since a uniform limb-darkened disk had been subtracted from the data, the background level of each scan is centered near 0 K. The y-axis indicates the relative increments in T$_b$. The separation between scans is 10 K. On the right side we indicate the frequency of each scan, in the same color as the scan (these numbers were rounded off to the lower value, i.e., 19 GHz corresponds to 19.5 GHz).
The scans are plotted as a function of planetographic latitude. The vertical dashed lines with a number at the top are lines through several zones and belts, as discussed in the text. At the top we show a slice through a visible light image of Jupiter from the OPAL program on HST, taken on 19 Jan. 2015 (Simon et al., 2015). 
}
\label{figscans}\end{figure*}

\begin{figure*}
\includegraphics[scale=0.7]{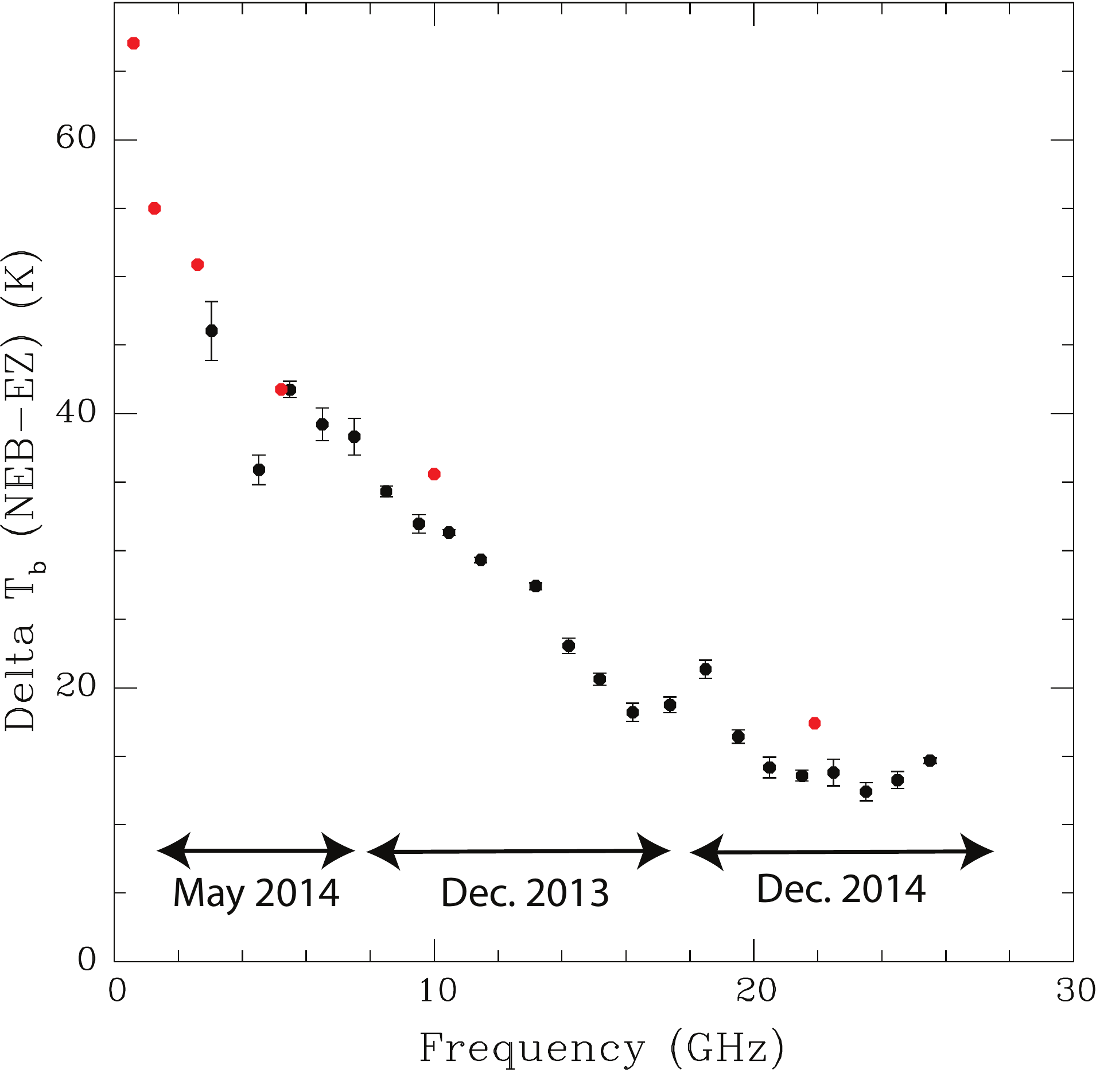}
\caption{Plot of the peak-to-peak difference in brightness temperature, $\Delta$ T$_b$ between the NEB and EZ as a function of frequency. The red dots are the values from the {\it Juno}/MWR PJ1 brightness temperatures (from Li et al., 2017).
}
\label{figdeltaTb}\end{figure*}

\begin{figure*}
\includegraphics[scale=0.6]{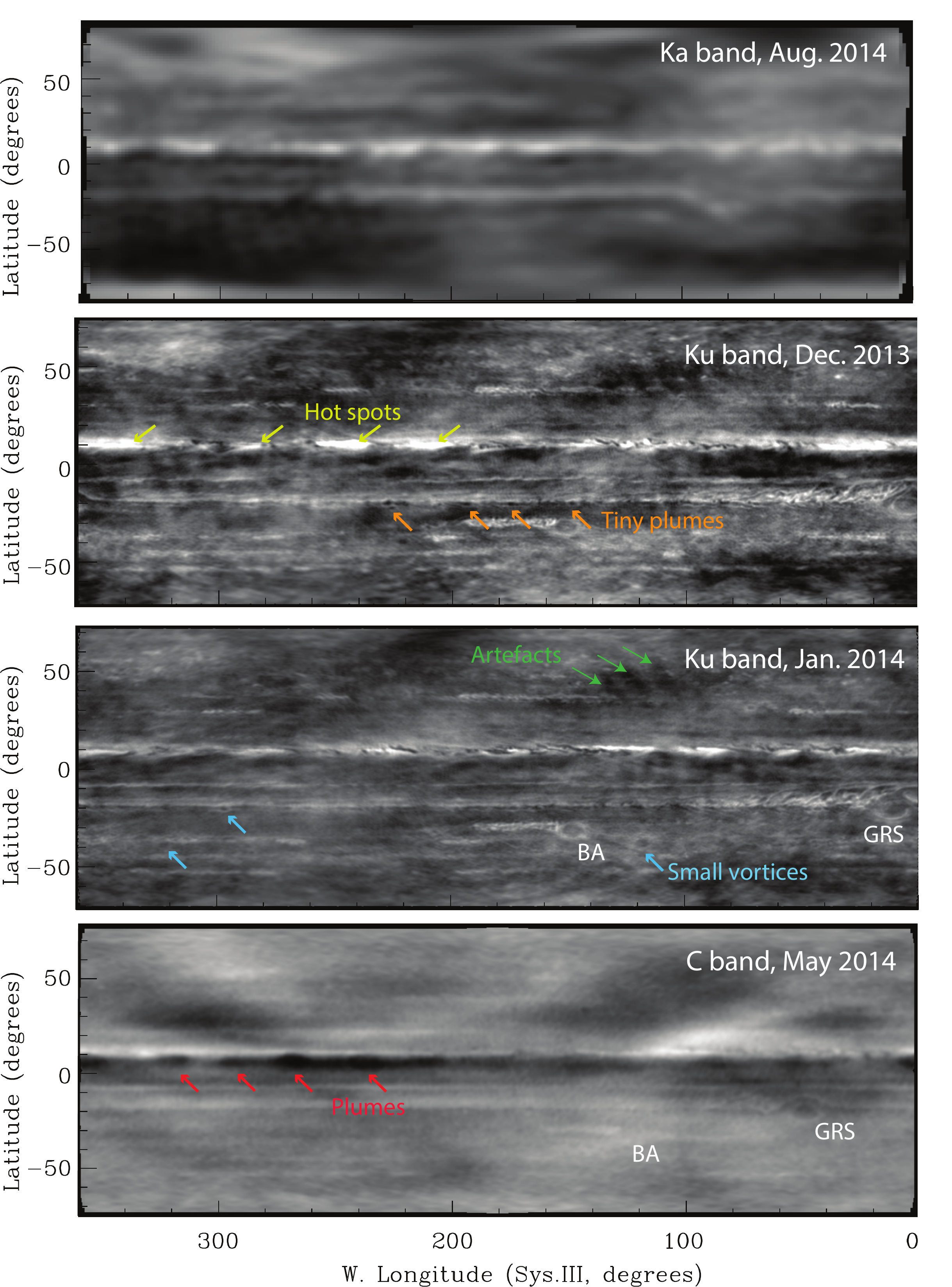}
\caption{a) Longitude-resolved maps at Ka (29--37 GHz), Ku 12--18 GHz), and C (4--8 GHz) bands, as indicated in the upper right hand corners. A few features discussed in the text are indicated: green arrows: large-scale artefacts; yellow: hot spots; red: ammonia plumes; orange: tiny plumes; cyan: small vortices.
}
\label{figresolved}
\end{figure*}

Longitude-resolved maps were presented in Section~\ref{resolved2}, Figure~\ref{figresolved}. As in Section~\ref{smeared3}, we ran chi-square fits to the data in combination with fits by eye to derive altitude profiles for the NH$_3$ abundance for several selected features. These fits are shown in Figure~\ref{figfeatures1} and the ammonia profiles are shown in Figure~\ref{figNH3profiles}b. These profiles may be slightly different from those published by dP16 due to a recalibration of the data (Section~\ref{obs}) and the addition of K band. We also used NH$_3$ abundances in the deep atmosphere that were smaller than in dP16; the values used here provide a better match to the C band (4--8 GHz) data. Our derived value ($4.1 \times 10^{-4}$, or 3.2 $\times$ solar N) is within the uncertainties of the values derived from the mass spectrometer Galileo Probe data (Wong et al., 2004a), and close to the deep atmospheric abundance determined from {\it Juno}/MWR data ($3.5 \times 10^{-4}$, Li et al. 2017). 
Results for Ka band are shown too, where available; but due to the lower spatial resolution in these data, the brightness temperatures should be considered either upper (plumes) or lower (hot spots, GRS) limits. Similarly, for any small features where we show minimum or maximum values these brightness temperatures should also be considered upper or lower limits, resp., simply because the convolution with our beam may have diluted the actual low and high values somewhat.
For comparison, as in Figs.~\ref{figmodelfits} and \ref{figmodelsjuno}, we show on all panels the results for the EZ (minimum T$_b$)and NEB (maximum T$_b$), at the appropriate emission angle (i.e., at the latitude of the feature considered). 

{\bf Hot spots} 

Hot spots are visible in the longitude-resolved maps (several are marked by yellow arrows in Fig.~\ref{figresolved}) at latitudes between $\sim$~8.5--11$^{\circ}$ N, forming the radio-hot belt on the south side of the NEB.  In order to get a realistic idea how deep we probe in hot spots, we determined the peak temperature in 5 different hot spots. These values were averaged, and the variation between them is shown as the standard deviation, or errorbar, on the figure. Note, though, that the brightness temperature in hot spots likely varies from feature to feature and over time, so that the errorbar is indicative of the spread in brightness temperature rather than an uncertainty in the values. The December and January data for the Ku and X bands are shown separately.  As shown, the hot spots overall are substantially depleted in ammonia gas. The NH$_3$ profile shows a gradual decrease from the deep atmosphere abundance at $\sim$~8 bar down to $\sim 10^{-5}$ in the NH$_3$ cloud layer ($\sim$~0.6 bar), and it stays subsaturated at higher atitudes (at RH$=$1\%). This profile is not unlike that observed by the Galileo Probe Net Flux Radiometer (NFR) (Sromovsky et al., 1998), though details differ somewhat as one might expect for features that vary substantially over time and across the planet.

{\bf Plumes} 

Large plumes, the most prominent features at C band (such as marked by red arrows on Fig.~\ref{figresolved}), extending across latitudes $\sim$~4--7.5$^{\circ}$ N. must cause (in part) the very low T$_b$ in longitude-smeared images (Fig.~\ref{figscans}). For a total of 9 plumes in each map we determined minimum brightness temperatures. As shown, the plumes are even colder than our nominal model. In addition to the NH$_3$ abundance profile for the plumes (black line in Fig.~\ref{figNH3profiles}b), we also show the nominal profile (dashed black line) to emphasize that the only way we can fit the K (18--26 GHz) and Ku (12--18 GHz) band data of the plumes is for NH$_3$ to be supersaturated up to the $\sim$~0.5 bar level; we also ``skipped'' the formation of the NH$_4$SH layer, i.e., the time scales for vertical ascent are shorter than that creating NH$_4$SH. This translates into plumes rising $\sim$~10 km above the nominal NH$_3$ condensation level. Such updrafts are expected to condense into fresh ammonia ice above the main NH$_3$-ice clouddeck. 
However, {\it Hubble} WFPC2 methane band images do not show enhancements of high-altitude haze particles associated with these features (Lii et al. 2010), as might be expected by strong dynamical uplift. To investigate the correlation between particle properties and NH$_3$ gas in these features, we obtained simultaneous VLA and {\it Hubble} observations in 2017; these data will be discussed in a future paper.

The degree of NH$_3$ supersaturation implied by our best-fit plume profile (solid black line in Fig.~\ref{figNH3profiles}b) is remarkable. In fact, in the 500-700 mbar range, the profile corresponds to several hundred percent supersaturation with respect to NH$_3$ ice. Such high supersaturation would be very different from the water cloud case in the terrestrial environment, where supersaturation is 10\% or less (Young 1993). We cannot explain the high NH$_3$ concentrations in the 500-700 mbar range in terms of a temperature anomaly (higher temperature allows higher concentrations before saturation is reached), because longitude-resolved temperature maps show no evidence for the 10 K anomalies that would be required (Fletcher et al. 2016). It is clear from the overall VLA dataset that these plumes have comparatively more NH$_3$ at all levels than any other parts of the planet. So even if there is an error in our quantitative determination of the NH$_3$ abundance in the plumes, it is clear qualitatively that the plume profiles of NH$_3$ are unique.

At the highest levels probed, in the K band, Figure~\ref{figresolved} shows that the plumes are among the radio-coldest areas of Jupiter, but they are joined by many other small cold features, particularly in turbulent areas. Brightness temperatures in small cold features in the GRS wake and Oval BA wake reach values as low as 130 K, just like in the plumes. These regions are consistent with ammonia being supplied from below, on a timescale faster than its photolysis lifetime of $\sim$~0.1 years (Edgington et al. 1998). The enriched ammonia gas in radio-dark regions in the K and Ku band are thus prime candidates for places to find spectrally-identifiable ammonia ice clouds (SIACs). Indeed, the most prominent SIACs reported by Baines et al. (2002) from NIMS data were in the NEB/EZ plumes and convective storms in the GRS wake, exactly where our data some years later find radio-cold features. Using {\it New Horizons} data, Reuter et al. (2007) reported a SIAC along the northern edge of what appears to be a large cyclonic vortex. The lack of concurrent high-resolution optical imaging precludes us from identifying cyclones in our 2013/2014 VLA dataset, but cyclones bear morphological similarities to the wakes of the GRS and Oval BA, where we do find radio-dark spots. Finally, longitudinally-averaged {\it Cassini}/CIRS data revealed SIACs at 23$^\circ$N, 2$^\circ$N, and a double peak at 9$^\circ$ and 13$^\circ$S (Fig. 11 in Wong et al. 2004b). In all these locations, our zonal means (Fig.~\ref{figscans}) show local minima in several of our 1-GHz wide frequency chunks (particularly K-band, line 3 in Fig.~\ref{figscans}). In fact, the strongest CIRS NH$_3$-ice signature at 23$^\circ$N is where we find the coldest zonal mean radio temperatures on Jupiter, at the 24 GHz frequency that probes the highest altitudes.

At the deepest levels, probed by C and S band maps, it is visually apparent from Fig.~\ref{figresolved} that the plumes (and the EZ) are the most NH$_3$-rich areas on the planet, and the hot spots (and the NEB) have the lowest NH$_3$ abundance. Dynamical flows are clearly still modulating the NH$_3$ abundance. Since NH$_3$-ice condensation is not an active loss process at depths of 1--10 bar,  flows must extend over the full range from $\sim$~0.5 bar down to over 20 bar. {\it In situ} measurements of water in the Galileo Probe entry site (Wong et al. 2004a) demonstrated that the Rossby-wave system producing 5-$\mu$m hot spots (and the radio plumes) reaches more than 20 bars. This is deep enough to mix material up from below the compositional barrier near 8.5 bar that we find at most latitudes in Fig.~\ref{figNH3}.

The plumes have an eastward phase velocity of 102.2$\pm$1.4 m/s relative to the System III coordinate system (dP16), very similar to the phase velocity determined for the hot spots when measured in 5-$\mu$m imaging (Ortiz et al., 1998). The fact that the radio plume and 5-$\mu$m hot spot velocities are essentially equal, bolstered the suggestion that the plumes are the deep signature of the equatorially trapped Rossby wave (dP16) that had been theorized to form the 5-$\mu$m hot spots (Showman and Dowling, 2000; Friedson, 2005). In an independent study, Fletcher et al (2016) detected these same plume and hot spot ``pairs'' in mid-infrared {\it Cassini} and IRTF/TEXES data, which they also interpreted as caused by Rossby waves.

{\bf GRS}

 We show brightness temperatures for the GRS that were median averaged over an area $\sim 4^\circ \times 2^\circ$, centered on the GRS. For the Ku and X band observations the numbers from December 2013 and January 2014 were averaged. The errorbar is the standard deviation over the area, and for Ku and X band the standard deviation from the two dates were added in quadrature. Since the GRS is relatively faint in C band, the measurement was taken from the 4-GHz wide map of the entire (4--8 GHz) C band. We similarly binned data in frequency in the 18--37 GHz range. As shown, the NH$_3$ abundance gradually decreases from the deep atmospheric abundance at and below 5 bar to 1.5$\times$10$^{-4}$ at 0.7 bar, and has a $\sim$~1\% humidity at higher altitudes. 

Our value essentially agrees with that determined from 5-$\mu$m spectroscopy by Bjoraker et al. (2018), who measured 2$\times$10$^{-4}$ over the range probed (0.7$<P<$5 bar). Also at mid-infrared wavelengths, probing $\sim$~0.5 bar level, NH$_3$ does not stand out in the GRS, in contrast to PH$_3$, which is considerably enhanced in the thermal-infrared (Fletcher et al., 2016; Bjoraker et al., 2018). {\it HST} observations in the ultraviolet (Edgington et al., 1999) show a NH$_3$ mixing ratio $\sim 8 \times 10^{-8}$ at the 250 mbar level, and Griffith et al. (1992) reported similarly a relative depletion in NH$_3$ over the GRS relative to the STZ by a factor of 4 from {\it Voyager} infrared spectroscopy. The latter authors also measured PH$_3$, and reported no difference in the PH$_3$ abundance over the GRS compared to the STZ. 
This is contrary to expectation, since one might have expected an enhancement in both NH$_3$ and PH$_3$ if the GRS represents a vortex with gas rising from the deep atmosphere. As pointed out by Griffith et al. (1992), at the higher altitudes NH$_3$ gas is subject to both condensation and photolysis, while PH$_3$ is only affected by photochemistry; near the tropopause, the chemical lifetimes for both species is similar to the time constant for vertical transport over an atmospheric scale height (Kay and Strobel, 1983). Since anticyclones are cold at the top (Fletcher et al., 2010; Marcus et al., 2013), perhaps condensation plays a larger role in limiting the NH$_3$ concentration than hitherto asssumed. 

As shown before, the ring on the south side of the GRS is bright at radio wavelengths. We used the maximum brightness temperature in this ring to get an indication of the NH$_3$ profile in this bright ring. The best fit NH$_3$ profiles are shown in Figure~\ref{figNH3profiles}b. 
The ring is depleted much more in NH$_3$ gas and to much deeper levels in the atmosphere than in the GRS itself. This agrees with observations at mid-infrared wavelengths (Fletcher et al., 2010).

{\bf Oval BA} 

As for the GRS, we show median averaged values of the brightness temperature in Oval BA, also averaged over  $\sim 4^\circ \times 2^\circ$, and the errorbars were determined in the same way as for the GRS. The oval was indistinguishable in the Ka band data, likely because of the low spatial resolution.
As shown, the NH$_3$ abundance drops from the nominal value to 1.75$\times$10$^{-4}$ at the $\sim$~8 bar level, and stays this low up through the NH$_3$ cloud condensation level; conditions are saturated from 700 mbar to 550 mbar, and humidity is 20\% at $P <$ 550 mbar, using our 
standard temperature profile. Groundbased mid-IR imaging showed that Oval BA is about 2 K colder than 
its surroundings, at the 400-mbar level (Cheng et al. 2008). Wong et al. (2011) argued that the thermal 
anomaly may not be fully resolved in the ground-based measurements, so that the actual temperature anomaly 
could be closer to 5 K colder than the surroundings. Accounting for this thermal anomaly, the NH$_3$ 
relative humidity at P $<$ 550 mbar in Oval BA may be 30-70\%, rather than 20\% with our nominal temperature 
profile.

{\bf Wakes of the GRS and Oval BA} 

The wakes of both the GRS and Oval BA are clearly extremely turbulent. In order to get an impression of the NH$_3$ abundances in the bright and dark regions, we plot the minimum and maximum brightness temperatures in both wakes, which represent maximum and minimum values for T$_b$, resp., due to beam convolution. A best-fit profile is shown (in red) for the maximum temperatures, which shows NH$_3$ abundances somewhat lower than in the GRS ring, but not as low as in the hot spots. The minimum brightness temperatures are compared with the nominal profile and with the supersaturated profile that was derived for the plumes. The values for the wake of Oval BA are superimposed on the nominal and the NEB profiles; there are too few points to derive a fit to these data. The errorbars were derived from the difference between the December 2013 and January 2014 data for the X and Ku bands.

{\bf Tiny plumes } 

Just north of Oval BA, at the south side of the SEB (at a latitude of 19$^\circ$ S), a series of small dark spots is apparent in Ku band and X band (marked with orange arrows in Fig.~\ref{figresolved}). We plotted the minimum brightness temperatures (i.e., upper limits to the actual brightness temperature) in these spots (3 in each of the December and January datasets) in Figure~\ref{figfeatures1}, with superposed the nominal and NEB models. These plots show that ammonia gas is also brought up in these tiny plumes. Judging from these and the maps in Figure~\ref{figresolved}, the numerous tiny dark spots in the SEB and NEB, as well as at other places on the planet, are probably all places where the full complement of NH$_3$ gas is brought up from the deep atmosphere. In several cases the dark spots are surrounded by bright rings, very similar to the small vortices seen at a wavelength of 5-$\mu$m (de Pater et al., 2010; 2011), indicative of dry subsiding air.

\section{Summary and Conclusions}\label{conclusion}

Longitude-smeared and longitude-resolved maps of Jupiter were created at frequencies from 3 up to 37 GHz from data obtained with the Very Large Array from December 2013 throughout 2014. The spatial resolution at disk center varies between $\sim$~1000 km up to $\sim$~4000 km. Numerous zones and belts can be distinguished at all frequencies between 7 and 25 GHz, while the NEB, SEB, and EZ are prominent even at the lowest (3 GHz) frequencies. In the longitude-resolved maps we see a multitude of large and small-scale features, such as ovals, including the GRS and Oval BA, radio-bright hot spots and radio-cold large and small plumes.
We summarize the following findings:

$\bullet$ {\it Radio-hot belt}: The radio-hot belt, extending over 8.5--11$^{\circ}$ N latitude, i.e., on the south side of the NEB near the interface with the EZ, is prominent at all frequencies. This band contains a series of hot spots, which make it stand out as a radio-hot belt.
Just to the south of this radio-hot belt are the ammonia plumes, which are particularly prominent at C band (4--8 GHz), probing down to the $\sim$~8-10 bar level. Although the S band data are somewhat compromised by synchrotron radiation, the plumes and radio-hot belt stand out.. Both C and S band data show that the NEB's latitudinal extent is increasing with decreasing frequency, confirming {\it Juno} results (Bolton et al., 2017; Li et al., 2017).

$\bullet$ {\it Weather layer}: In addition to the GRS, Oval BA, and the large oval-shaped ammonia plumes, a multitude of small-scale features is visible at frequencies $\gtrsim$~8 GHz, and essentially absent at lower frequencies, which formed the foundation of previous conclusions (dP16) that there is much small-scale dynamics in the ``weather'' layer of Jupiter's atmosphere, at $P<$2--3 bar, i.e., that region in the atmosphere where clouds form. 

$\bullet$ {\it NH$_3$ latitude vs altitude distribution}: Assuming all variations in brightness temperature are caused by spatial (3D) variations in the ammonia abundance, we derived a longitude-averaged map of NH$_3$ as a function of latitude and altitude (pressure), roughly at pressures between $\sim$~0.5 and 10--20 bar. 
At all latitudes, the NH$_3$ profile is consistent with one that is either constant or decreasing with altitude. The formation of the NH$_4$SH layer is apparent at many latitudes between $\pm 50^\circ$ (N and S), while at higher latitudes the atmosphere between $\sim$~1 and 10 bar is characterized by a relatively low NH$_3$ abundance ($\sim 1.75 \times 10^{-4}$). 

$\bullet$ {\it EZ}: As shown before (dP16; Bolton et al., 2017; Li et al., 2017), the ammonia abundance in the EZ is equal to that in Jupiter's deep atmosphere ($\sim 4 \times 10^{-4}$ at $P>$10--20 bar), which is near the lower limit detected by the Galileo probe ($3.5 \times 10^{-4}<{\rm NH_3}<7.8 \times 10^{-4}$) based on mass spectrometer data (Wong et al. 2004a), but smaller than the deep (15-bar) concentration of $6.35 \pm 0.9 \times 10^{-4}$ from the probe's radio signal attenuation (Hanley et al. 2009, Folkner et al. 1998).
Our data are consistent with gas rising up from the deep atmosphere, with some ammonia being lost in the solution cloud (6--7 bar), by forming the NH$_4$SH layer at $\sim$~2.5 bar and the ammonia ice cloud at $\sim$~0.7--0.8 bar. Above the NH$_3$ ice cloud, using our nominal TP profile, NH$_3$ is at a relative humidity of 50\%, likely caused by a combination of photolysis (at P $<$ 0.3 bar) and dynamics.

$\bullet$ {\it NEB}: Ammonia gas is depleted in the NEB compared to the deep atmosphere down to at least the 20 bar level; its mixing ratio is $1.75 \times 10^{-4}$. At altitudes above $\sim$~1.5 bar the abundance is decreasing down to $10^{-5}$ in the NH$_3$-ice cloud, and it is subsaturated (RH$=$1\%) at higher altitudes.

$\bullet$ {\it GRS \& Oval BA}: The ammonia abundance profile in the GRS, decreasing from $\sim$~4 $\times$ 10$^{-4}$ at $\sim$~5 bar to 1.5 $\times$ 10$^{-4}$ at 0.7 bar with a $\sim$~1\% relative humidity above, is consistent with values derived at 5-$\mu$m (Bjoraker et al., 2018) and in the mid-infrared (Fletcher et al., 2016). Since anticyclones are cold at the top, condensation may play a larger role in limiting the NH$_3$ concentration than hitherto asssumed. Oval BA has a NH$_3$ abundance of $\sim$~1.75 $\times$ 10$^{-4}$ throughout the upper atmosphere (0.7--8 bar).

$\bullet$ {\it Hot spots \& plumes}: 
The NH$_3$ abundance in the hot spots shows  a gradual decrease from the deep atmospheric value at $\sim$~8 bar down to $10^{-5}$ at 0.6 bar, with a low (RH$=$1\%) humidity above.  In contrast, the large ammonia plumes just south of the hot spots show a clear supersaturation of NH$_3$ gas up to 0.5 bar, i.e., NH$_3$ gas rises $\sim$~10 km above the main NH$_3$ cloud deck before it condenses out. The hot spots and plumes together are signatures of the equatorially trapped Rossby wave. 

Numerous tiny NH$_3$ plumes are visible scattered across Jupiter's disk, in particular in the SEB at 19$^\circ$ S and in the radio-hot belt. Radiative transfer calculations for these tiny plumes are consistent with NH$_3$ gas being brought up from the deep atmosphere before condensing at high altitudes.

$\bullet$ {\it Comparison with Juno}: Our RT models that best fit the longitude-averaged VLA data at 4--25 GHz match the {\it Juno}/MWR PJ1 data extremely well, despite the fact that the VLA data were averaged over longitude and taken two years before the {\it Juno} PJ1 data, and the latter data were taken along one north-south track at frequencies between 0.6 and 22 GHz. The difference in NH$_3$ profiles derived from the {\it Juno} data by Bolton et al. (2017) and Li et al. (2017) with the profile derived by us from the VLA data may be attributed to differences in the TP and NH$_3$ profiles within and above the NH$_3$ cloud layer. Since the brightness temperature as observed is an integrated quantity (i.e., integrated over a range of altitudes),  differences at these altitudes do affect the modeled brightness temperatures over the 10--25 GHz range. In the future we will  derive the NH$_3$ profile from a combined {\it Juno}--VLA dataset, using VLA data taken simultaneously with {\it Juno} observations. We will also combine such an analysis with simultaneously obtained mid-infrared data to differentiate between spatial variations in temperature and ammonia gas, i.e., break the degeneracy between these quantities, as discussed in Section~\ref{CIRS} of this paper.  

\section*{Appendix}

Figure~\ref{figbeams} shows an example “beam pattern” image for the longitude-resolved maps in Fig.~\ref{figresolved}.
As noted earlier, the longitude-resolved maps are formed by stitching together a number of facets. Each dot in Fig.~\ref{figbeams}  is centered on a facet and represents the point-spread function at that location. The resolution achieved varies between facets, particularly with latitude.
 
For different observing bands, the different frequency and array configurations used will also lead to quite different resolutions. The set of facets used was also varied between observing bands.

\section*{Acknowledgements}

We thank Cheng Li for providing us with the {\it Juno} PJ1 nadir brightness temperatures of Jupiter as published in Li et al. (2017), as well as the TP profile used in their analysis. 

This research was supported by NASA Planetary Astronomy (PAST) awards NNX14AJ43G and 80NSSC18K1001, and NASA Outer Planets Research Program award NNX11AM55G to the University of California, Berkeley. Fletcher was supported by a Royal Society Research Fellowship and European Research Council Consolidated Grant at the University of Leicester. The National Radio Astronomy Observatory (NRAO) is a facility of NSF operated under cooperative agreement by Associated Universities, Inc. VLA data used in this report, associated with project code 13B-064, are available from the NRAO Science Data Archive at https://archive.nrao.edu/archive/advquery.jsp. 

\pagebreak
\section{References}

Achterberg, R.K., Conrath, B.J., Gierasch, P.J., 2006. Cassini CIRS retrievals of ammonia in Jupiter's upper troposphere. 
Icarus 182, 169–180. doi: 10.1016/j.icarus. 2005.12.020. 

Ackerman, A.S., Marley, M.S., 2001. Precipitating condensation clouds in substellar atmospheres. Astrophys. J. 556, 872-884.

Asplund, M., Grevesse, N., Sauval, A.J., Scott, P., 2009. The chemical composition of
the Sun. Ann. Rev. Astron. Astrophys. 47, 481-522.

Atreya, S.K., 1986, Atmospheres and ionospheres of the outer planets and 
their satellites, Springer-Verlag, Berlin Heidelberg New York London Paris 
Tokyo

Atreya, S.K., Romani, P.N., 1985. Photochemistry and clouds of Jupiter, Saturn and Uranus. In Planetary Meteorology (G. E. Hunt, Ed.), pp. 17–68. University Press, Cambridge.

Baines, K.H.,  Carlson, R.W, Kamp, L.W., 2002. Fresh Ammonia Ice Clouds in Jupiter. I.
Spectroscopic Identification, Spatial Distribution, and Dynamical Implications. Icarus
159, 74-94.

Bellotti, A., Steffes, P.G., Chinsomboom, G., 2016. Laboratory measurements of the 5-20 cm wavelength opacity of ammonia, water vapor, and methane under simulated conditions for the deep jovian atmosphere. Icarus, 280, 255-267.

Berge, G.L., Gulkis, S., 1976. Earth based radio observations of 
Jupiter: millimeter to meter wavelengths, 1976,
in Jupiter, ed. T.Gehrels, p. 621-692, Univ. of 
Arizona Press, Tucson.

Bjoraker, G.L., Wong, M. H., de Pater, I., Hewagama, T., \'Ad\'amkovics, M., Orton, G.S., 2018. The Gas Composition and Deep Cloud Structure of Jupiter’s Great Red Spot. Astron. J., 156, \#101, 15 pp..

Bolton, S.J., et al., 2017. Jupiter's interior and deep atmosphere:The initial pole-to-pole passes with the Juno spacecraft. Science, 356, 821-825.

Burke, B.F., Franklin, K.L., 1955. Observations of a variable radio source 
associated with the planet Jupiter, J. Geophys. Res., 60, 213-217.

Butler, B. J., Steffes, P. G., Suleiman, S. H., Kolodner,
    M. A., Jenkins, J. M., 2001. Accurate and Consistent Microwave
    Observations of Venus and Their Implications. Icarus, 154, 226-238.

Cheng, A.F., and 14 colleagues, 2008. Changing characteristics of Jupiter’s Little Red Spot. Astron. J. 135, 2446–2452.

de Pater, I., 1986. Jupiter's zone-belt structure at radio wavelengths: 
  II. Comparison of observations with model atmosphere calculations, 
Icarus,  68, 344-365.

de Pater, I., Dickel, J.R., 1986. Jupiter's zone-belt structure at radio
  wavelengths: I. Observations, Astrophys. J., 308, 459-471.

de Pater, I., Massie, S.T., 1985: Models of the millimeter-centimeter 
  spectra of the Giant planets, Icarus,  62, 143-171.

de Pater, I., Dunn, D.E., 2003.  VLA Observations of Jupiter's
Synchrotron Radiation at 15 and 22 GHz. Icarus, 163, 449-455.

de Pater, I., Kenderdine, S., Dickel, J.R., 1982. A new interpretation 
  of Jupiter's thermal disk based upon radio data at 6, 11 and 21 cm,
  Icarus, 51, 25-38.

de Pater, I., Schulz, M., Brecht, S. H., 1997.
Synchrotron evidence for Amalthea's influence on Jupiter's
electron radiation belt,  J. Geoph. Res.,  102, No. A10,
22,043 - 22,064.

de Pater, I., Dunn, D.,  Zahnle, K.,  Romani, P.N.,
2001. Comparison of Galileo Probe Data with Ground-based Radio
Measurements. Icarus, 149, 66-78.

de Pater, I., Butler, B., Green, D.A.,  Strom, R., Millan, R., Klein, M.J.,
Bird, M.K., Funke, O., Neidhofer, J., Maddalena, R., Sault, R.J.,
Kesteven, M., Smits, D.P., Hunstead, R., 2003. Jupiter's radio
spectrum from 74 MHz up to 8 GHz. Icarus, 163, 434-448.

de Pater, I., DeBoer, D.R., Marley, M., Freedman, R.,  Young, R.,
2005. Retrieval of water in Jupiter's deep atmosphere using
microwave spectra of its brightness temperature. Icarus, 173, 425-438.

de Pater, I., Wong, M. H., Marcus, P. S., Luszcz-Cook, S., \'Ad\'amkovics, M., 
Conrad, A., Asay-Davis, X., Go, C., 2010.
   Persistent Rings in and around Jupiter's Anticyclones - Observations and Theory.
Icarus, 210, 742-762.

de Pater, I., Wong, M. H., de Kleer, K., Hammel, H. B., \'Ad\'amkovics, M.,
Conrad, A., 2011. 
Keck Adaptive Optics Images of Jupiter's North Polar Cap and Northern Red Oval. Icarus, 213, 559-563. 

de Pater, I.,  Fletcher, L.N.,  Luszcz-Cook, S.H., DeBoer, D.,
    Butler, B., Hammel, H.B., Sitko, M.L., Orton, G.,  Marcus, P.S., 2014. Neptune's Global
Circulation deduced from Multi-Wavelength Observations. Icarus,
    237, 211-238.

de Pater, I., Sault, R. J., Butler, B., DeBoer, D., Wong, M. H., 2016. Peering through Jupiter's Clouds with Radio Spectral Imaging, Science, 352, Issue 6290, pp. 1198-1201. (referred to as dP16).
 
Devaraj, K., Steffes, P.G., Duong, D., 2014. The centimeter-wavelength opacity of ammonia under deep jovian conditions. Icarus, 241, p. 165-179.

Edgington, S. G., Atreya, S. K., Trafton, L.M., Caldwell, J.J., Beebe, R.F., Simon, A.A., West, R.A.,  Barnet, C., 1998. On the Latitude Variation of Ammonia, Acetylene, and Phosphine Altitude Profiles on Jupiter from HST Faint Object Spectrograph Observations.  Icarus 133,  192--209.  

Edgington, S. G., Atreya, S. K., Trafton, L.M., Caldwell, J.J., Beebe, R.F., Simon, A.A., West, R.A.,  1999. Ammonia and Eddy Mixing Variations in the Upper Troposphere of Jupiter from HST Faint Object Spectrograph Observations.  Icarus 142,  342--356.  

Fletcher, L.N., Orton G.S., Yanamandra-Fisher, P., Fisher, B.M.,
 Parrish, P.D., Irwin, P.G.J., 2009. Retrievals of atmospheric
 variables on the gas giants from ground-based mid-infrared
 imaging. Icarus, 200, 154-175.

Fletcher, L.N. \& 13 colleagues, 2010. Thermal structure and composition of Jupiter's
Great Red Spot from high-resolution thermal imaging. Icarus 208, 306–318.

Fletcher, L.N., Baines, K.H., Momary, T.W., Showman, A.P., Irwn, P.G.J., Orton, G.S., Roos-Serote, M., Merlet, C., 2011. Saturn’s tropospheric composition and clouds from Cassini/VIMS 4.6–5.1 lm nightside spectroscopy. Icarus, 214, 510-533.

Fletcher, L. N., Greathouse, T. K., Orton, G. S., Sinclair, J. A., Giles, R. S., Irwin, P. G. J., Encrenaz, T., 2016. Mid-infrared mapping of Jupiter's temperatures, aerosol opacity and chemical distributions with IRTF/TEXES. Icarus, 278, p. 128-161. 

Friedson, A.J., 2005. Water, ammonia, and H$_2$S mixing ratios in Jupiter’s five-micron hot spots. A dynamical model. Icarus 177, 1-17.

Folkner, W.M., Woo, R., Nandi, S., 1998. Ammonia abundance in Jupiter's atmosphere derived from the attenuation of the Galileo Probe's radio signal. J. Geophys. Res. 103, 22847-22855.

Gibson, J., Welch, Wm. J.,  de Pater, I., 2005. Accurate Jovian
Flux Measurements at $\lambda$1cm Show Ammonia to be Sub-saturated in
the Upper Atmosphere. Icarus, 173, 439-446.

Griffith, C. A., B\'ezard, B., Owen, T.,  Gautier, D., 1992. The tropospheric abundances of NH3 and PH3 in Jupiter’s Great Red Spot from Voyager IRIS observations. Icarus 98, 82–93.

Hanley, T.R., Stefes, P.G., Karpowicz, B.M., 2009. A new model of the hydrogen and helium-broadened microwave opacity of ammonia based on extensive laboratory measurements. Icarus, 202, 316-335. doi:10.1016/j.icarus.2009.02.002

Ingersoll, A.P., Gierasch, P.J., Banﬁeld, D., Vasavada, A.R., the Galileo
imaging team, 2000. Moist convection as an energy source for the large-scale motions in Jupiter’s atmosphere. Nature 403, 630–632.

Janssen, M. J., et al., 2017. MWR: Microwave Radiometer for the Juno Mission to Jupiter. Space Sci. Rev. 213, 139-185. DOI 10.1007/s11214-017-0349-5.

Karim, R. L., deBoer, D., de Pater, I., Keating, G. K., 2018. A Wideband Self-consistent Disk-averaged Spectrum of Jupiter Near 30 GHz and Its Implications for NH$_3$ Saturation in the Upper Troposphere. Astron. J., 155, article id. 129, 8 pp.

Karpowicz, B. M.,  Steffes, P. G., 2011.
In search of water vapor on Jupiter: Laboratory measurements of the
microwave properties of water vapor under simulated jovian
conditions. Icarus, 212, 210-223.

Kaye, J. A.,  Strobel, D.F., 1983. HCN formation on Jupiter: The coupled photochemistry of ammonia and acetylene. Icarus 54, 417-433.

 Klein, M. J.,  Gulkis, S., 1978, Jupiter's atmosphere: Observations and 
interpretation of the microwave spectrum near 1.25 cm wavelength, Icarus, 35, 44-60.

Larson, H. P., Davis, D. S., Hofmann, R., Bjoraker, G. L., 1984. The Jovian atmospheric window at 2.7 $\mu$m: A search for H$_2$S. Icarus, 60, 621-639.

Li, C., et al., 2017. The distribution of ammonia on Jupiter from a preliminary
inversion of Juno microwave radiometer data. Geophys. Res. Lett., 44, 5317–5325, doi:10.1002/2017L073159.

Lii, P.S., Wong, M.H., de Pater, I., 2010. Temporal variation of the tropospheric cloud and haze in the jovian equatorial zone. Icarus 209, 591-601.

Lindal, G.F., 1992, The atmosphere of Neptune: An analysis of radio 
occultation data acquired with Voyager 2, Astron. J., 103, 967-982.

Marcus, P.S.,  Asay-Davis, X.,  Wong, M. H., de Pater, I.,
2013. Jupiter's New Red Oval: Dynamics, Color, and Relationship to
Jovian Climate Change.  Journal of Heat Transfer, 135, 011007-1 to  011007-9; DOI: 10.1115/1.4007666

Orton, G.S., Gustafsson, M., Burgdorf, M., Meadows, V., 2007. Revised ab initio
models for H2-H2 collision-induced absorption at low temperatures. Icarus 189,
544–549.

Marcus, P.S., Tollefson, J., Wong, M.H., de Pater, I., 2019. An equatorial thermal wind equation: Applications to Jupiter. Icarus, In Press.

Mayer, C.H., McCullough, T.P., Sloanaker, R.M., 1958. Observations of 
Mars and Jupiter at a wavelength of 3.15 cm, Astrophys.J., 127, 11-16.

Moeckel, C., Janssen, M.J., de Pater, I., 2018. A fresh look at the Jovian radio emission as seen by
Cassini-RADAR and implications for the time
variability. Icarus, submitted.

Ortiz, J.L., Orton, G.S., Friedson, A.J., Stewart, S.T., Fisher, B.M., Spencer, R.J.,
1998. Evolution and persistence of 5-μm hot spots at the Galileo probe
entry latitude. J. Geophys. Res. 103, 23051–23069.

Perley, R. A., Chandler, C. J., Butler, B. J., Wrobel, J. M.,
    2011. The Expanded Very Large Array: A New Telescope for New
    Science. Astrophys. J. Lett., 739, article id. L1, 5 pp.

Perley, R.A., Butler, B.J., 2013. An Accurate Flux Density
    Scale from 1 to 50 GHz. Astrophys. J. Suppl. 204, Issue 2, article id. 19, 20 pp.

Perley, R.A., Butler, B.J., 2017. An Accurate Flux Density Scale from 50 MHz to 50 GHz. Astrophys. J. Suppl., 230, Issue 1, article id. 7, 18 pp. 

Radhakrishnan, V., Roberts, J.A., 1960. Polarization and angular extent 
of the 960 Mc/sec radiation from Jupiter, Phys. Rev. Lett., 4, 493-494.

Reuter, D. C.,  Simon-Miller, A.A.,  Lunsford, A., Baines,  K. H., Cheng, A. F., Jennings,  D. E., Olkin,  C. B., Spencer, J. R., Stern, S. A., Weaver, H. A., Young, L.A., 2007 Jupiter Cloud Composition, Stratification, Convection, and Wave Motion: A View from New Horizons.  Science 318,  223-225.

Sault, R. J., Teuben, P. J., \& Wright, M. C. H. 1995, A Retrospective View of MIRIAD, in ASP Conf. Ser. 77,
Astronomical Data Analysis Software and Systems IV, ed. R. A. Shaw,
H. E. Payne, \& J. J. E. Hayes (San Francisco: ASP), 433-436.

Sault, R.J., Engel, C., de Pater, I., 2004. Longitude-resolved Imaging
of Jupiter at $\lambda=2$ cm. Icarus, 168, 336-343.

Showman, A.P., de Pater, I., 2005. Dynamical implications of Jupiter's
tropospheric ammonia abundance. Icarus, 174, 192-204.

Showman, A.P.,  Dowling, T.E., 2000. Nonlinear Simulations of Jupiter’s 5-Micron Hot Spots.
Science 289, 1737-1740.

Simon, A .A . , Wong, M.H. , Orton, G.S. , 2015. First 
results from the Hubble OPAL program: Jupiter in 
2015. Astrophys. J. 812 (1), 55.

Sromovsky, L.A., Fry, P.M., 2010. The source of 3-$\mu$m absorption in Jupiter's clouds: Reanalysis of ISO observations using new NH$_3$ absorption models. Icarus, 210, 211-229.

Sromovsky, L.A., Fry, P.M., 2010. The source of widespread 3-$\mu$m absorption in Jupiter's clouds: Constraints from 2000 Cassini VIMS observations. Icarus, 210, 230-257.

Sromovsky, L.A., Collard, A.D., Fry, P.M., Orton, G.S., Lemmon, M.T.,
Tomasko, M.G., Freedman, R.S., 1998.  Galileo Probe measurements of
thermal and solar radiation fluxes in the jovian atmosphere. 
J. Geophys. Res., 103, 22929--22978.

Tollefson, J., de Pater, I., Marcus, P.S., Luszcz-Cook, S., Sromovsky, L.A., Fry, P.M., Fletcher, L.N., Wong, M.H., 2018. Vertical wind shear in Neptune's upper atmosphere explained with a modified thermal wind equation. Icarus, 311, 317-339.

Weidenschilling, S.J., Lewis, J.S., 1973. Atmospheric and cloud structures of the jovian planets. Icarus 20, 465-476.

Weiland, J.L., et al., 2011. Seven-year Wilkinson Microwave
    Anisotropy Probe (WMAP) observations: Planets and celestial
    calibration sources. Astrophys. J. Suppl. 192: 19 (pp.21)

Wong, M.H., Mahaffy, P.R., Atreya, S.K., Niemann, H.B., Owen,  T.C., 2004a. Updated Galileo probe mass spectrometer measurements of carbon, oxygen, nitrogen, and sulfur on Jupiter. Icarus 171, 153-170. 

Wong, M. H., Bjoraker, G. L., Smith,  M. D., Flasar, F. M., Nixon, C.A., 2004b. Identification of the 10-$\mu$m ammonia ice feature on Jupiter.  Planetary and Space Science 52,  385--395.

Wong, M.H., de Pater, I., Asay-Davis, X.S., Marcus, P.S., Go, C.Y. , 2011. Vertical structure of Jupiter's 
  Oval BA before and after it reddened: What changed? Icarus 215, 211-225.

Wong, M. H., Atreya, S. K., Kuhn, W. R., Romani, P. N., \& Mihalka, K. M.
2015. Fresh clouds: A parameterized updraft method for calculating
cloud densities in one-dimensional models. Icarus, 245, 273

Young, K.C. (1993) Microphysical processes in clouds. Oxford University Press, New York NY USA.

\pagebreak

\setcounter{figure}{7}

\begin{figure*}
\includegraphics[scale=0.6]{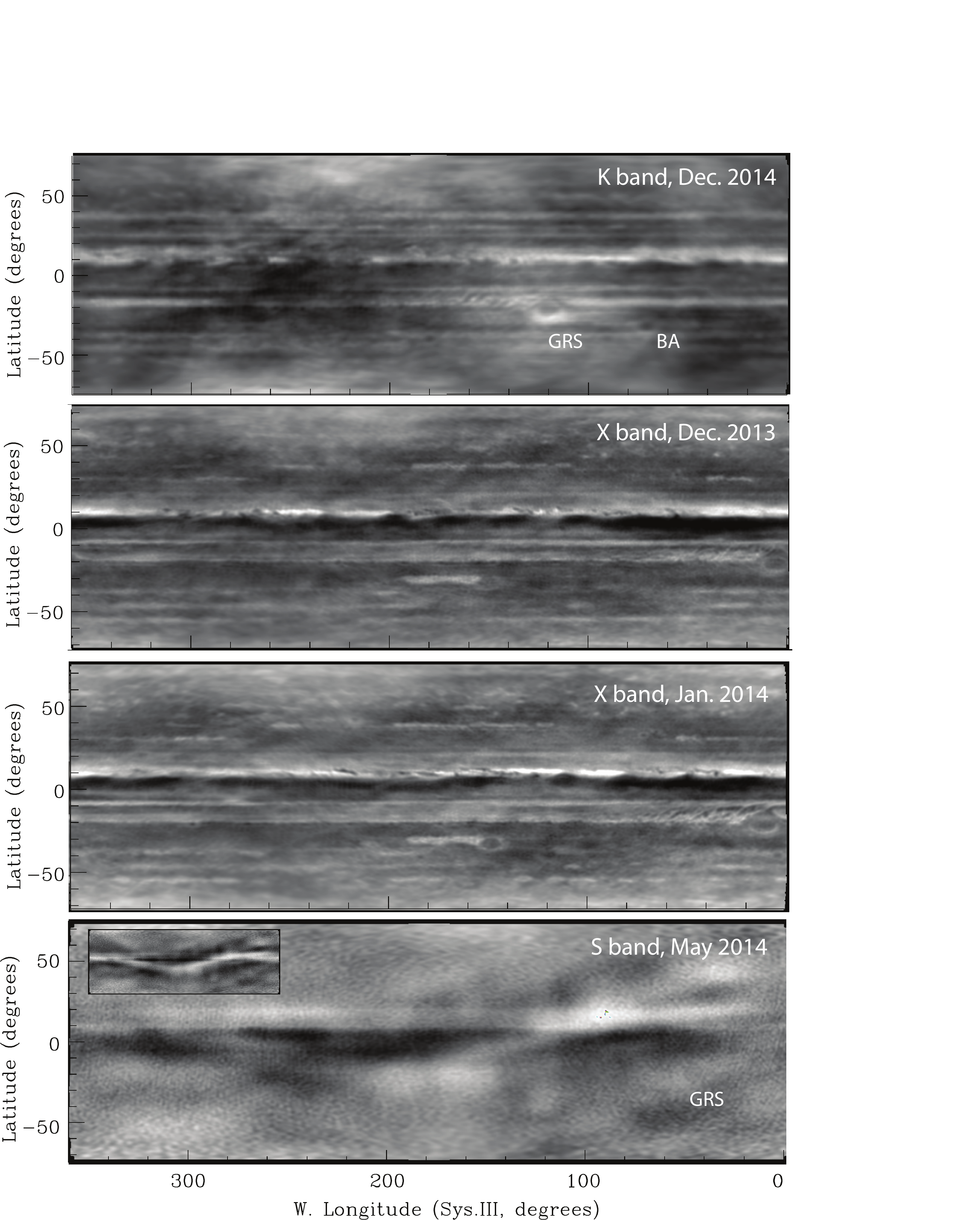}
\caption{b) Longitude-resolved maps at K (18--26 GHz), X (8--12 GHz), and S (2--4 GHz)  bands. 
}
\label{figresolvedb}
\end{figure*}

\begin{figure*}
\includegraphics[scale=0.80]{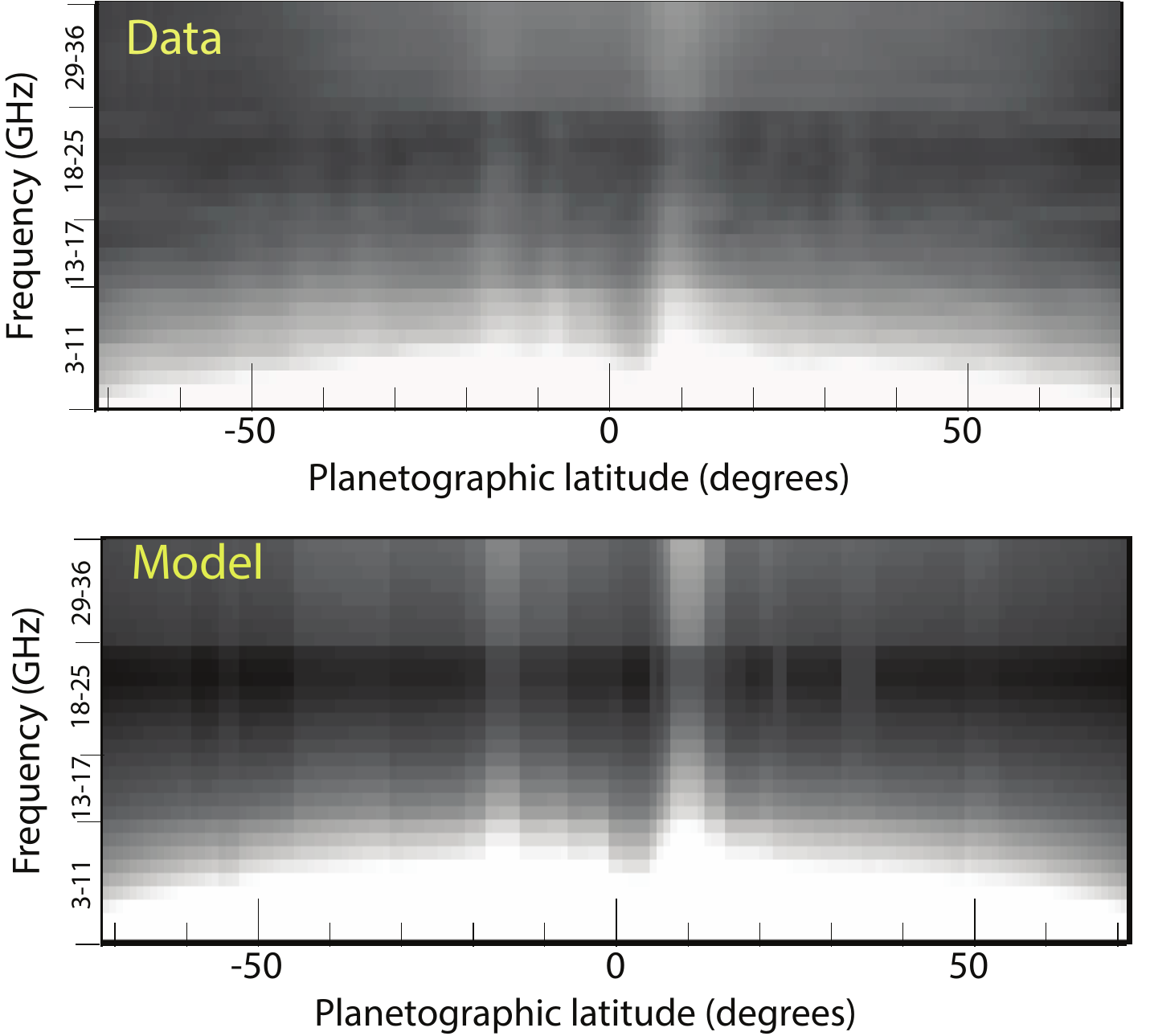}
\caption{top: A map of the scans displayed in Fig.~\ref{figscans}, with frequency along the y-axis and planetographic longitude along the x-axis.
bottom: A map of the models that best fit the data. The brightness scale in the two panels was optimized to bring out the variety of zones and belts, and are slightly different for the two panels. 
}
\label{figSCXUK}\end{figure*}


\begin{figure*}
\includegraphics[scale=0.70]{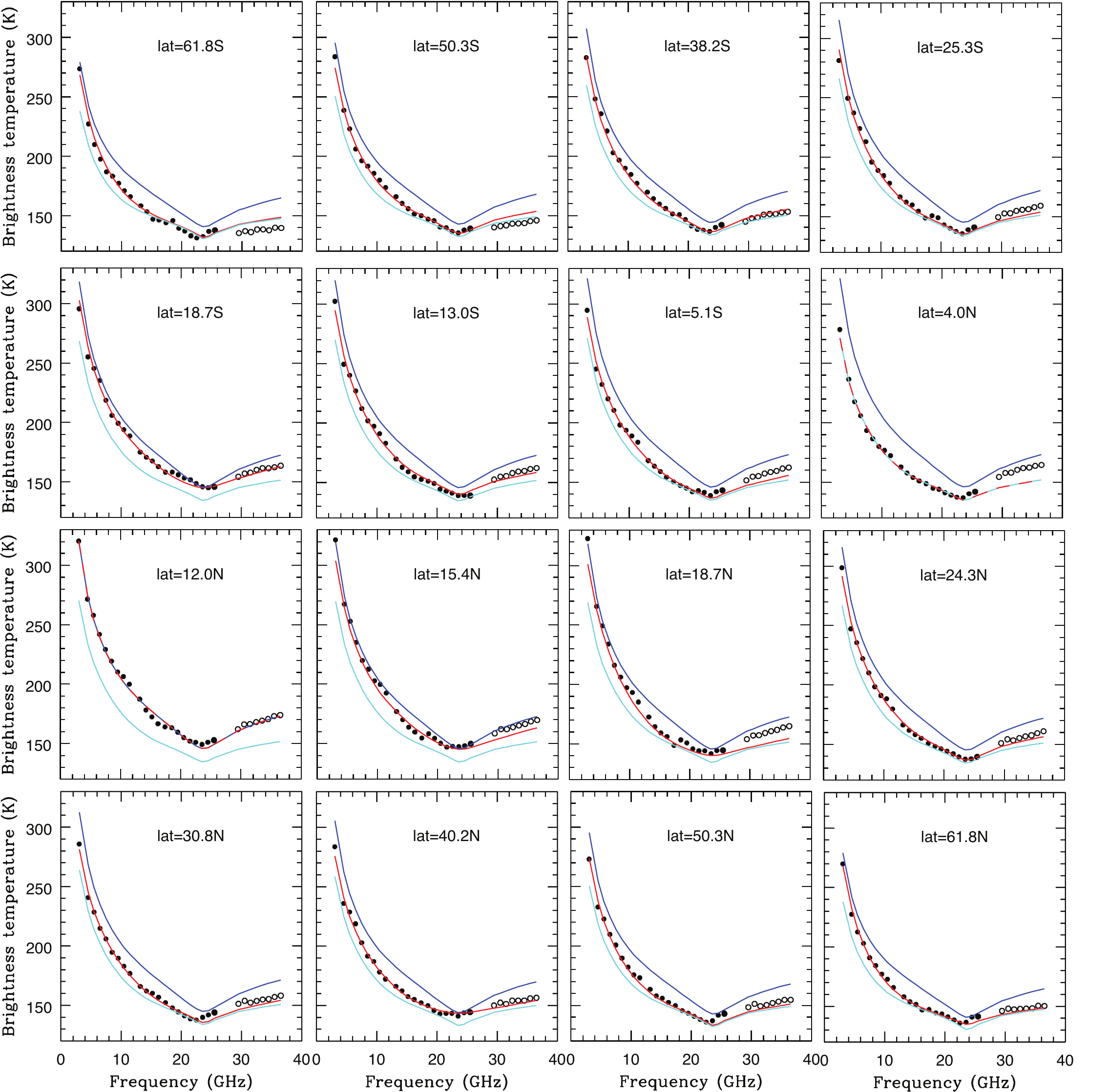}
\caption{Longitude-averaged brightness temperatures T$_b$ (black dots) and superposed best-fit model spectra (red lines) for several latitudes. The ammonia abundances for the modeled spectra are shown in Fig.~\ref{figNH3}. The data at 29-37 GHz (open circles) were not used to constrain the fits, but are shown for comparison with the model spectra. These points were omitted from the fits because they have a 4 $\times$ lower spatial resolution. The 3 GHz data point was not used in the fits either, due to possible contamination by Jupiter's synchrotron radiation. The cyan and blue curves show the models with the parameters (NH$_3$ profile) that best fit the EZ and NEB (the red curve coincides with the cyan at 4.0N, and with the blue one at 12.0N latitude).
}
\label{figmodelfits}\end{figure*}


\begin{figure*}
\includegraphics[scale=0.70]{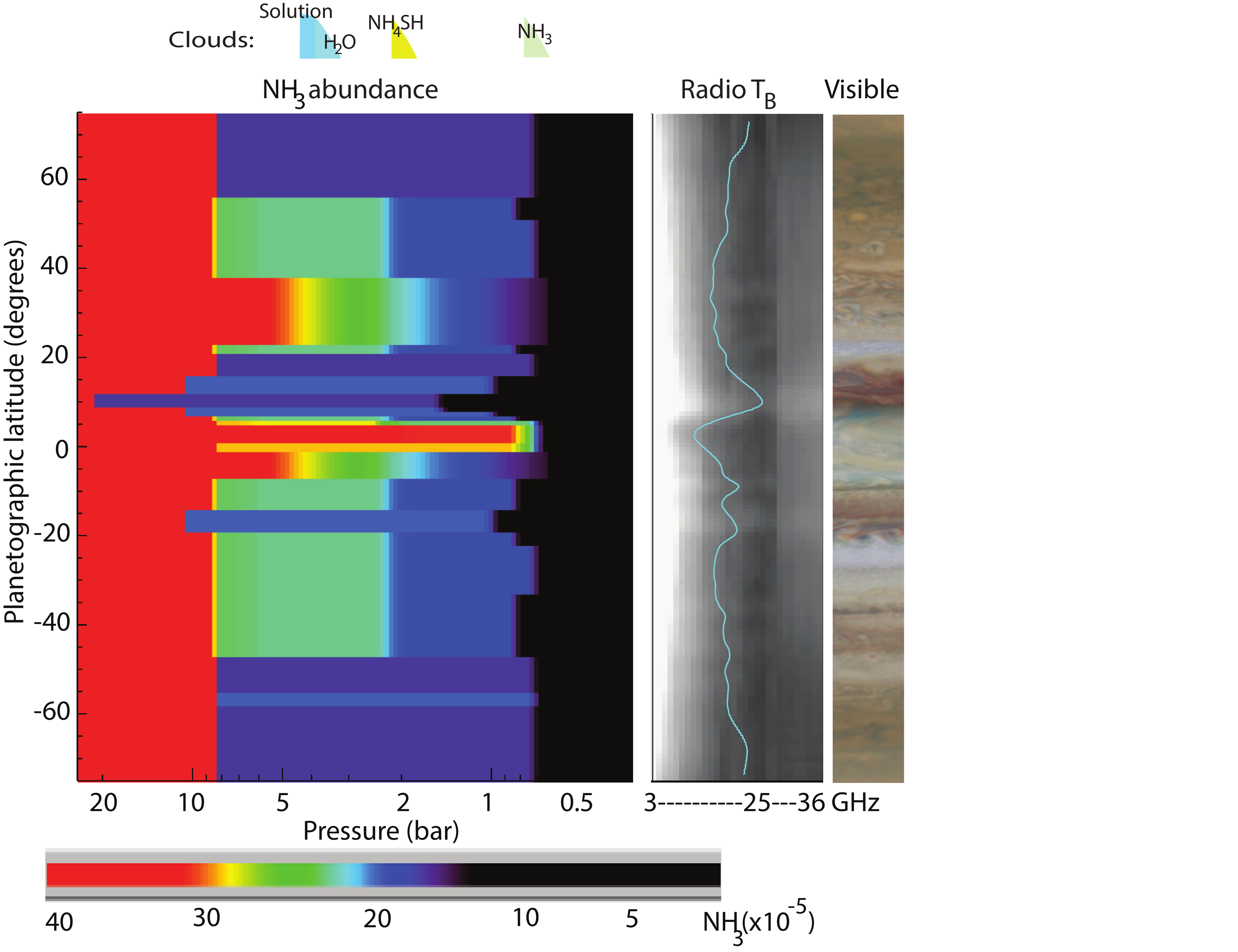}
\caption{Ammonia abundance as obtained by fitting the VLA data is shown on the left, with a scalebar at the bottom. The approximate location of the various cloud layers is shown at the top. The image in the middle shows the brightness temperature map from Fig~\ref{figSCXUK}, with superposed the 10 GHz brightness temperature scan from Fig.~\ref{figscans}. On the right the same slice through Jan. 2015 OPAL data is shown as at the top of Fig.~\ref{figscans} (from Simon et al., 2015). 
}
\label{figNH3}\end{figure*}


\begin{figure*}
\includegraphics[scale=0.50]{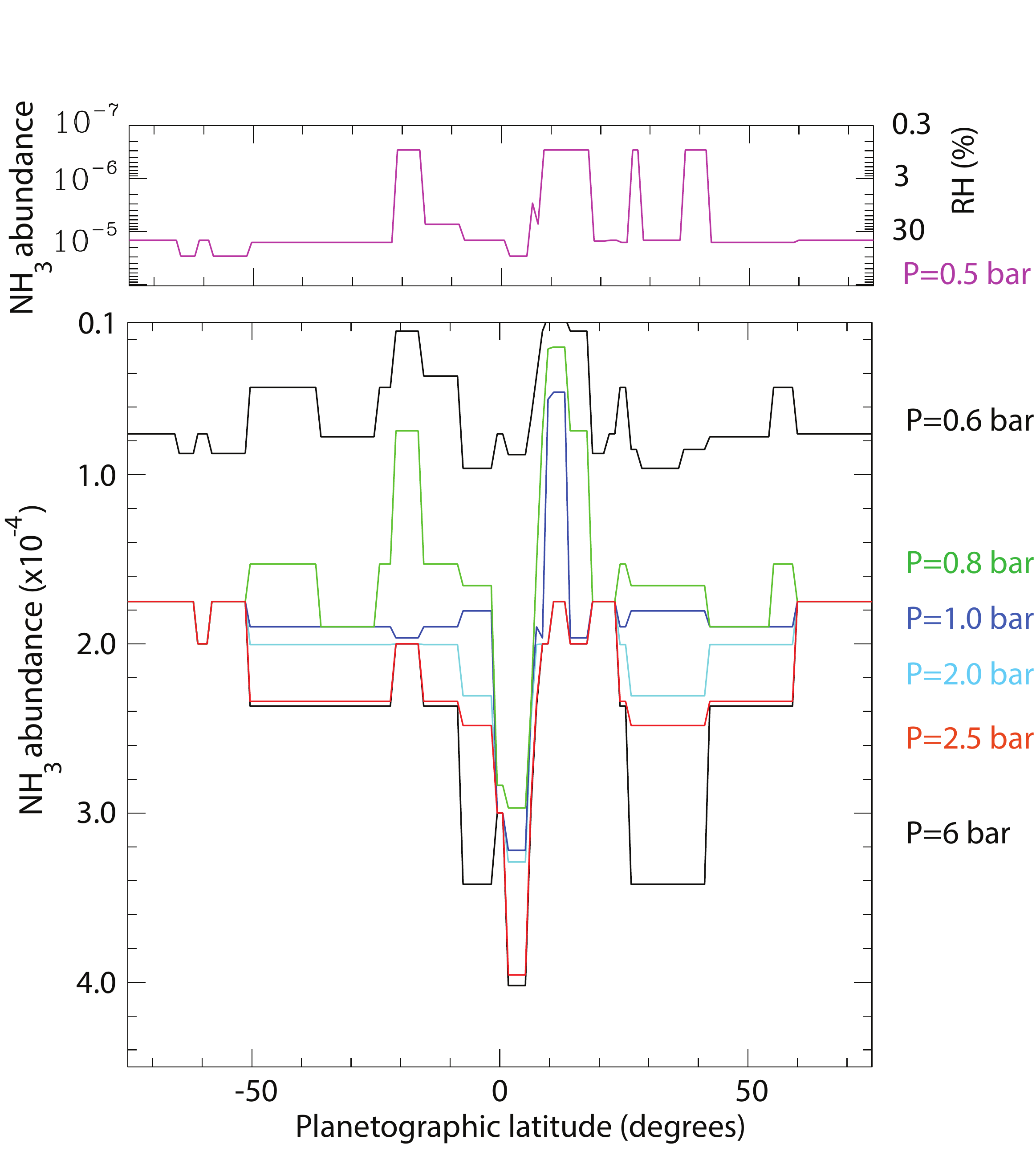}
\caption{Slices at several pressure levels between 6 and 0.5 bar through the NH$_3$ abundance map shown in Fig.~\ref{figNH3}. These show quantitatively what the abundance map shows in color. 
}
\label{figNH3slice}
\end{figure*}


\begin{figure*}
\includegraphics[scale=0.80]{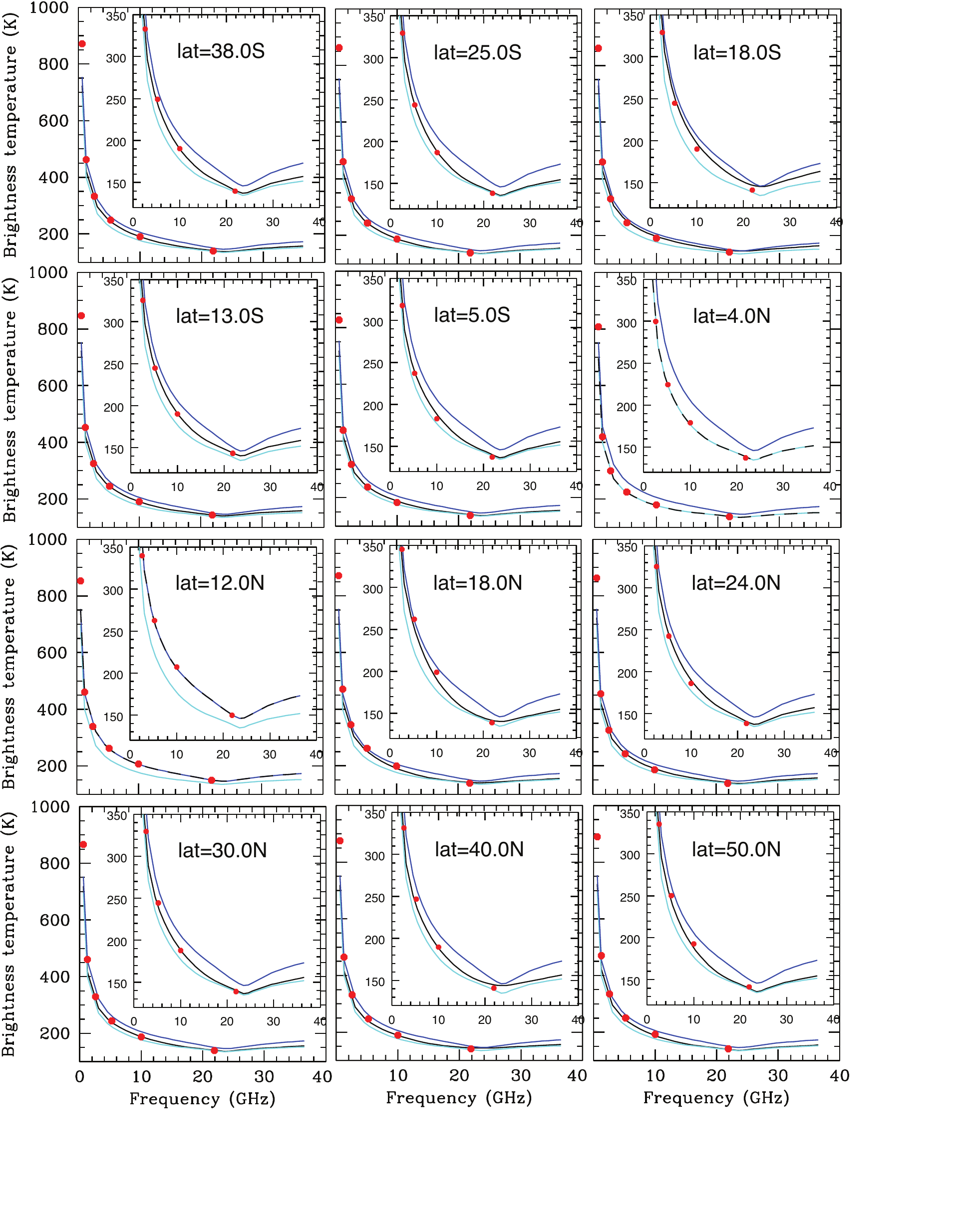}
\caption{Our best fit models to VLA data (black lines) superposed on the PJ1 Juno data (red dots). The inset shows the models $+$ data on the same scale as in Fig.~\ref{figmodelfits}. The Juno data (and superposed models) are for nadir viewing. The data are taken from Li et al. (2017). Blue and cyan synthetic spectra are based on our NEB and EZ models, respectively.
}
\label{figmodelsjuno}\end{figure*}

\begin{figure*}
\includegraphics[scale=0.50]{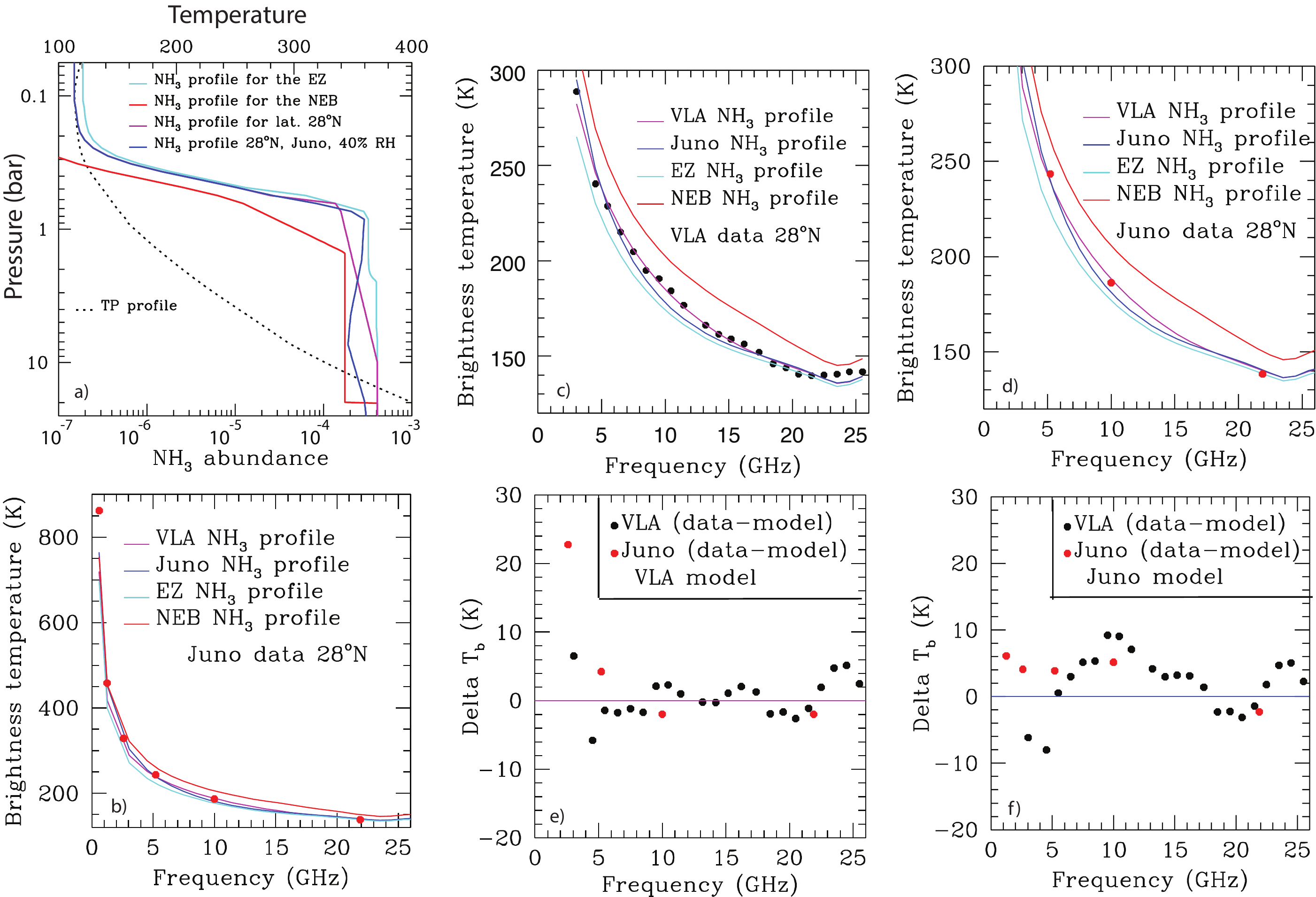}
\caption{a) Various NH$_3$ profiles as used in subsequent panels, with the same color scheme. The cyan and red curves are the NH$_3$ profiles that give a best match to the EZ and NEB spectra as measured with the VLA. The magenta curve is the profile that best matches the VLA data at a latitude of 28$^\circ$. The blue curve is the NH$_3$ profile that was derived by the {\it Juno} team to best fit the {\it Juno} data at a latitude of 28$^\circ$ (scanned in from Li et al., 2017). b) Spectra based upon the NH$_3$ profiles in panel a superposed on the {\it Juno} data at 28$^\circ$ latitude; all at nadir viewing. c) Spectra using the profiles in panel a superposed on the VLA data at 28$^\circ$ latitude. d) Zoomed in version from panel b. e) Difference in brightness temperature for the VLA and {\it Juno} data as compared to the VLA (magenta) model. f) Difference in brightness temperature for the VLA and {\it Juno} data as compared to the {\it Juno} (blue) model. Note that the difference in brightness temperature was taken from panel c for the VLA data, and from panel d for the {\it Juno} data.
}
\label{figjunoprofile}\end{figure*}


\begin{figure*}
\includegraphics[scale=0.62]{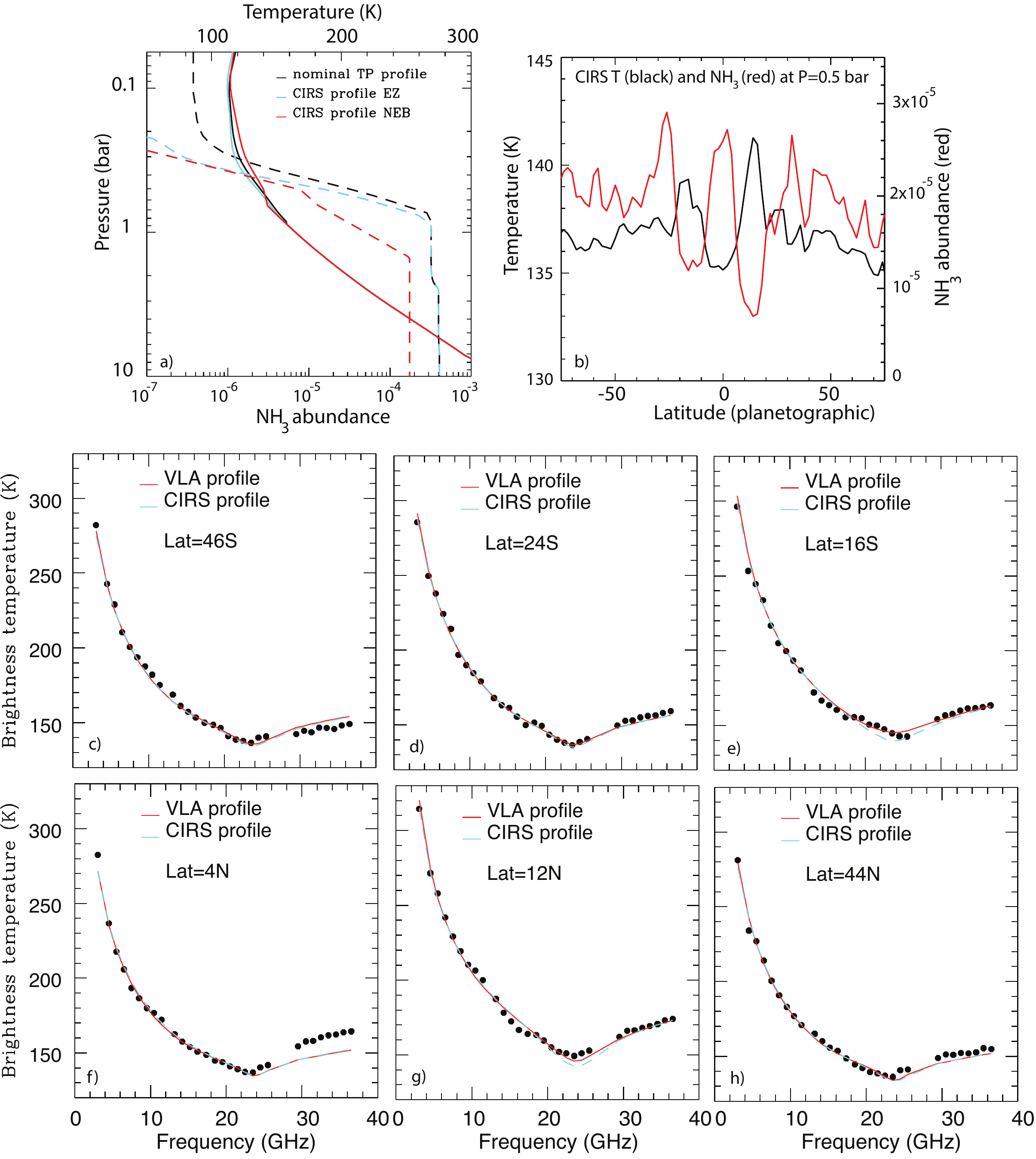}
\caption{a) TP profiles (solid lines) and NH$_3$ abundances (dashed lines): the black lines are our nominal profiles; cyan and red lines are for the EZ and NEB, where the TP and NH$_3$ values at $P \lesssim$ 0.6 were replaced by the CIRS-derived values. b) CIRS temperature (black) and NH$_3$ abundance (red) at a pressure of 0.5 bar. c-h) Spectra based upon our best fit models (in red) and based upon profiles where the TP and NH$_3$ values at $P \lesssim 0.6$ bar were replaced by the CIRS profiles from Fletcher et al. (2016).
}
\label{figmodelsCIRS}\end{figure*}


\begin{figure*}
\includegraphics[scale=0.50]{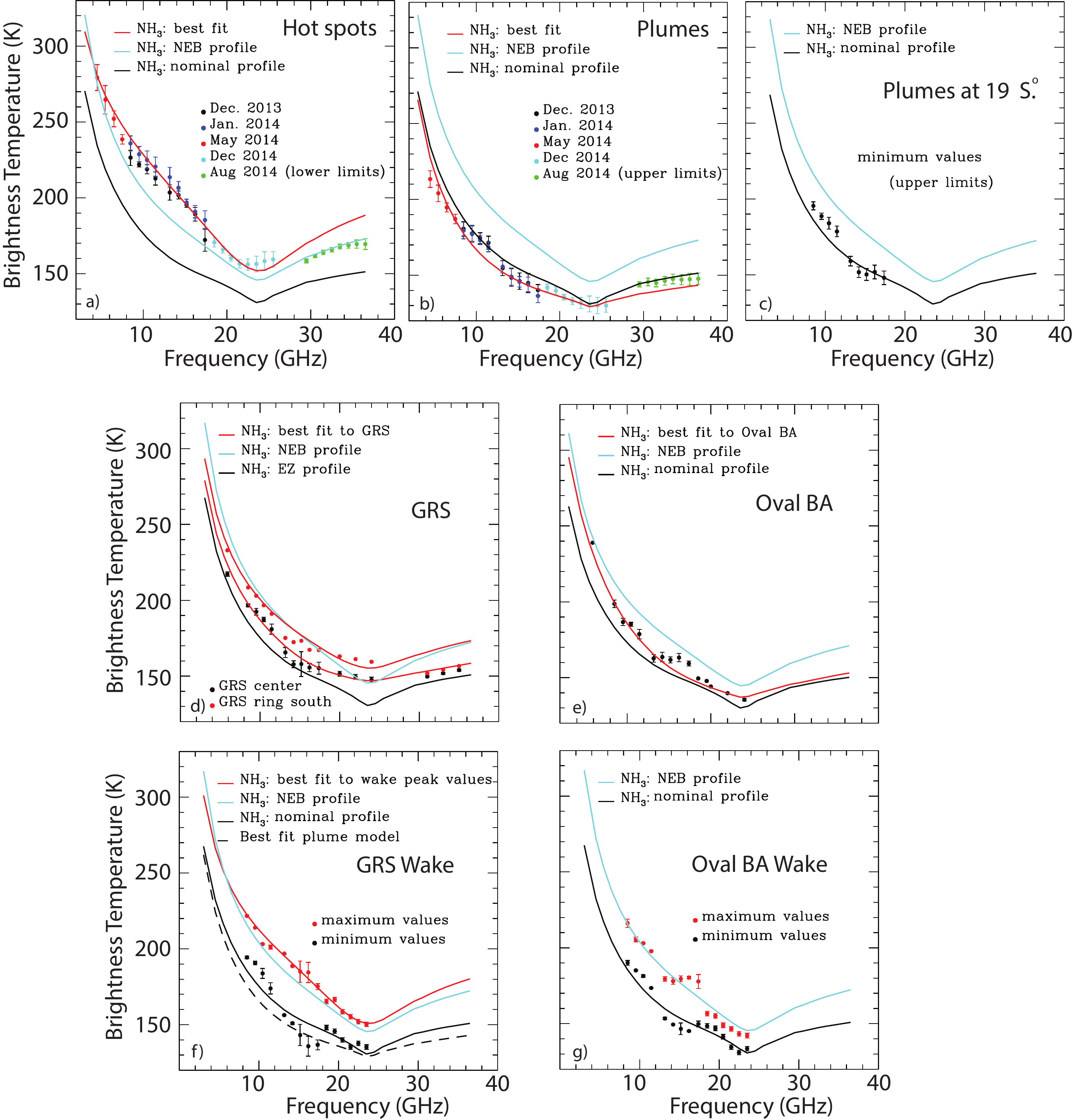}
\caption{Model fits superposed on data for several individual features. Best fits are shown in red. In panel d two red lines are shown: the lower one is for the center of the GRS, the upper one for the bright ring surrounding the GRS. The ammonia abundances for these models are shown in Fig.~\ref{figNH3profiles}. The data at 29-37 GHz were not used in the fit as these data have a $\sim$~4 times lower spatial resolution. 
}
\label{figfeatures1}\end{figure*}


\begin{figure*}
\includegraphics[scale=0.50]{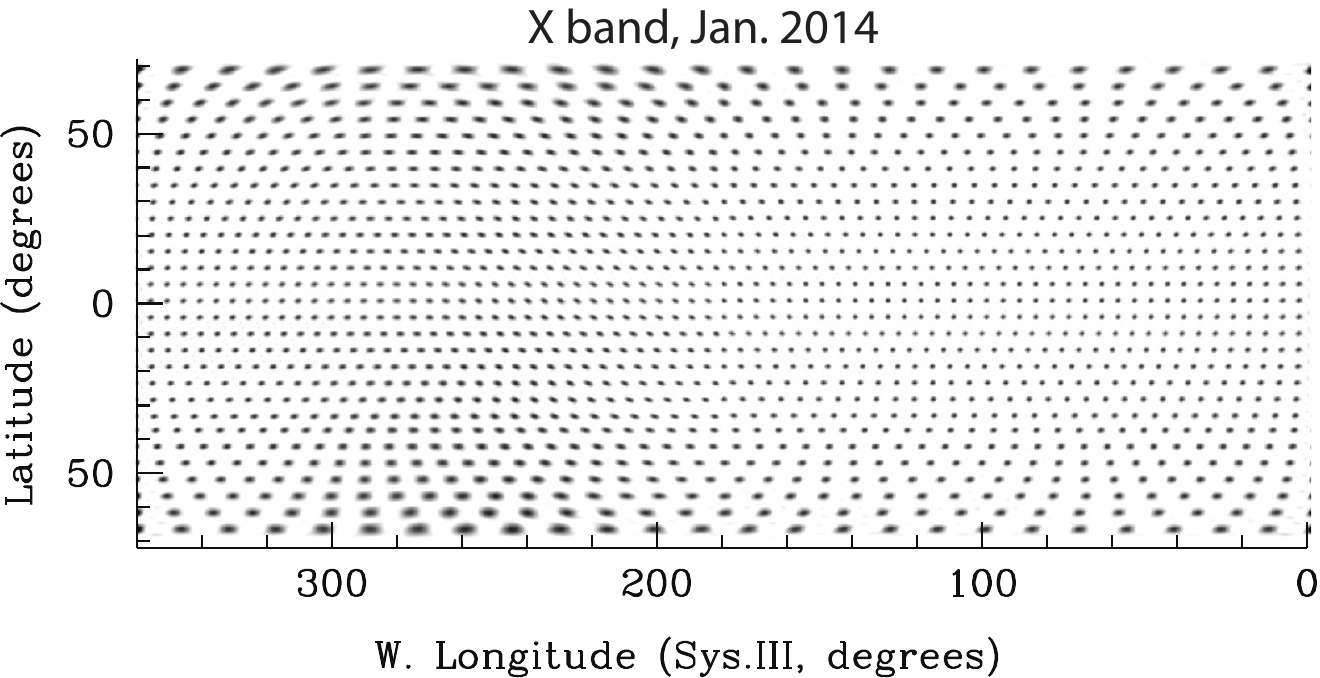}
\caption{Beam pattern of the longitude-resolved X-band map in Fig.~\ref{figresolved}.
}
\label{figbeams}\end{figure*}

\begin{table*}[!ht]  
\small
\caption{VLA Observations}
\vspace{0.15in}
\begin{tabular}{l c l c  c c c c c c c c  }
\hline
\hline
Date (UT)& VLA Array & Band & Frequency & Wavelength &$\Delta$& R (equ) & R (pol)& Obs-lat  \\
year/month/day& configuration&  &range (GHz) & center (cm) &(AU)  & (\pr) & (\pr)& (\degr)& \\
\hline
2013/12/23 & B & Ku& 12-18&2& 4.236 & 23.272 & 21.762 & 1.85 \\
2013/12/23 & B & X & 8-12 & 3& 4.236 & 23.272 & 21.762 & 1.85 \\
2014/01/09 & B &Ku& 12-18&2 &4.214 & 23.394 & 21.877  & 1.87 \\
2014/01/09 & B & X & 8-12 & 3& 4.214 & 23.394 & 21.877  & 1.87 \\
2014/01/09 & B & L & 1-2  & 20 &4.214  & 23.394 & 21.877 & 1.87 \\
2014/05/04 & A & C & 4-8 &5 & 5.637 & 17.487 &16.353 & 1.35 \\
2014/05/04 & A & S & 2-4 &10 & 5.637 & 17.487 &16.353 & 1.35 \\
2014/05/04 & A & L & 1-2 &20 & 5.637 & 17.487 &16.353 & 1.35 \\
2014/08/16 & D & Ku&12-18&2&  6.237&15.803 &14.778 & 0.82 \\
2014/08/16 & D & K& 18-26 &1.4& 6.237&15.803 &14.778 & 0.82 \\
2014/08/16 & D & Ka&29-37&0.9 & 6.237&15.803 &14.778 & 0.82 \\
2014/12/27 & C & K & 18-26 &1.4&  4.594 &21.458 & 20.066 &-0.26 \\ 
2014/12/27 & C & X & 8-12 &3&  4.594 &21.458 & 20.066 &-0.26 \\ 
2014/12/27 & C & C & 4-8 &5& 4.594 &21.458 & 20.066 &-0.26 \\ 
\\
\hline
\end{tabular}\label{tab.1}

$\Delta=$geocentric distance, R $=$ radius (equator and polar), and
Obs-lat is the observer's (or sub-) latitude.

\end{table*}


\small

\begin{table*}[!ht]  
\small
\caption{Disk-averaged brightness temperatures of the VLA maps}
\begin{tabular}{l c l c c c c c c c c c c c  }
\hline
\hline
Array &Band & Center Frequ & T$_b$(peak) & $q$ & Size &T$_b$(av)& T$_{\rm res}$ & T$_{\rm cmb}$ & T$_{\rm synch}$&T$_b$(map)&T$_b$(UV)&
T$_b$(final) \\
configuration& & (GHz)  & (K) &   &scaling  &(K) & (K) & (K) &(K) & (K)\\
\hline
A& S & 3.04    &260&0.2&1.00&237.1&-1.32&2.7&24& 238.5$\pm$7.2 & &(270)\\
A& C & 4.52    &248.5&0.16&1.001&230.7&-0.16&2.62&9&233.1& &233.1$\pm$7\\
A& C & 5.49    &236.4&0.16&1.001&219.4&0.09&2.60&5.5&222.0& &222.0$\pm$6.6\\
A& C & 6.50    &223.8&0.16&1.001&207.7&0.59&2.58&3.4&210.9& &210.9$\pm$6.3\\
A& C & 7.50    &211.3&0.16&1.001&196.1&0.52&2.55&2.3&199.2& &199.2$\pm$6.0\\
B& X & 8.50     &194.12&0.16&1.0012&180.2&0.62&2.53&1.6&183.4&  &188.9$\pm$5.7\\
B& X & 9.52     &188.3&0.16&1.0012&174.8&0.45&2.50&1.2&177.7& &183.0$\pm$5.5\\
B& X & 10.46   &182.9&0.16&1.0012&169.8&0.59&2.48&0.9&172.8& &178.0$\pm$5.3\\
B& X & 11.46   &177.1&0.16&1.0012&164.4&0.77&2.46&0.7&167.6& &172.6$\pm$5.2\\
B$+$D& Ku & 13.18 &160.9&0.08&1.000&155.0&9.0&2.42&0.5 &166.0&162.0 &164.0$\pm$4.9\\
B$+$D& Ku & 14.21 &158.1&0.08&1.000&152.3&5.6&2.40&0.4 &159.9&157.5 &158.7$\pm$4.8\\
B$+$D& Ku & 15.18 &154.4&0.08&1.000&148.7&4.9&2.38&0.3 &155.6&154.4 &155.0$\pm$4.7\\
B$+$D& Ku & 16.21 &150.7&0.08&1.000&145.1&4.7&2.36&0.3 &151.9&150.6 &151.3$\pm$4.5\\
B$+$D& Ku & 17.38 &146.1&0.06&1.000&142.1&5.4&2.33&0.2 &149.6&147.4 &148.5$\pm$4.5\\
C$+$D& K  & 18.49 &142.0&0.065&1.000&137.7&9.0&2.31&0.20&148.8&144.6 &146.7$\pm$4.4\\
C$+$D& K  & 19.51 &140.0&0.065&1.000&135.3&7.5&2.28&0.17&144.9&142.7 &143.8$\pm$4.3\\
C$+$D& K  & 20.49 &138.0&0.065&1.000&133.8&4.2&2.26&0.15&140.1&140.0 &140.1$\pm$4.2\\
C$+$D& K  & 21.51 &136.0&0.065&1.000&131.8&3.8&2.24&0.13&137.7&137.8 &137.8$\pm$4.1\\
C$+$D& K  & 22.49 &134.0&0.065&1.000&130.0&3.5&2.22&0.11&135.6&136.4 &136.0$\pm$4.1\\
C$+$D& K  & 23.51 &132.0&0.065&1.000&128.0&3.9&2.20&0.10&134.0&135.9 &135.0$\pm$4.1\\
C$+$D& K  & 24.49 &133.0&0.065&1.000&128.9&7.0&2.18&0.09&138.1&137.2 &137.7$\pm$4.1\\
C$+$D& K  & 25.51 &137.0&0.065&1.000&132.9&3.3&2.16&0.08&138.3&139.7 &139.0$\pm$4.2\\
D         &Ka & 29.49 &149.5&0.100&1.000& 142.7&1.5&2.08&0&146.2& &146.2$\pm$4.4\\
D         &Ka & 30.51 &150.7&0.101&1.000& 143.7&1.6&2.05&0&148.3& &148.3$\pm$4.4\\
D         &Ka & 31.49 &151.8&0.102&1.000& 144.7&1.7&2.04&0&148.4& &148.4$\pm$4.5\\
D         &Ka & 32.51 &153.0&0.103&1.000& 145.7&2.1&2.02&0&149.8& &149.8$\pm$4.5\\
D         &Ka & 33.49 &154.1&0.1045&1.000&146.6&1.8&2.00&0&150.4& &150.4$\pm$4.5\\
D         &Ka & 34.51 &155.2&0.106&1.000& 147.6&0.9&1.98&0&150.5& &150.5$\pm$4.5\\
D         &Ka & 35.49 &156.3&0.1075&1.000&148.6&1.3&1.96&0&151.8& &151.8$\pm$4.6\\
D         &Ka & 36.51 &157.5&0.109&1.000& 148.6&1.0&1.94&0&152.5& &152.5$\pm$4.6\\
\hline
\end{tabular}\label{tab.S1}

All data were scaled to the 2013 Dec. 23 epoch, with  Requ=23.272 and Rpol=21.7621

$^*$: These numbers were scaled to match the spectrum as observed at neighboring wavelengths. 

Column 4: Peak brightness temperature of the limb-darkened disk that gave a best fit to the u-v data.

Column 5: Limb-darkening parameter of the limb-darkened disk that fits the u-v data.

Column 6: Scaling factor of the fitted disk.

Column 7: Disk-averaged temperature of the limb-darkened disk that was subtracted.

Column 8: Residual disk-averaged brightness temperature, after subtraction of the limb-darkened disk. 

Column 9: Cosmic microwave background correction, T$_{\rm cmb}$.

Column 10: Disk-averaged brightness temperature due to Jupiter's synchroton radiation.

Column 11: Final disk-averaged brightness temperature based upon the maps: T$_b$(final)= T$_b$(av)+T$_{\rm res}$+T$_{\rm cmb}$-T$_{\rm synch}$. We adopted a 3\% calibration error.
For the C, S, and X band data, where only high resolution configurations had been used (A, B), no correction for synchrotron radiation was made, since this radiation was essentially resolved out.

Column 12: Best fit uniform disk to the u-v data of the low resolution configurations, at spacings $<$ 20 kilo-lambda. The T$_{\rm cmb}$ (Col 9) had been added, and the synchrotron radiation subtracted (see text).

Column 13: Final disk-averaged brightness temperature. Since the X band data were low by 3\%, we multiplied the brightness temperatures in Column 11 by 1.03, analogous to the treatment by de Pater et al. (2016) where all data at X and C band were multiplied by a factor of 1.065. The value of 270 K at S band has been estimated from the disk-averaged spectrum; it is not shown in Fig. 4.

\end{table*}
%
\small

\begin{table*}[!ht]  
\small
\caption{Details on the longitude-resolved VLA maps}
\vspace{0.15in}
\begin{tabular}{l c l c  c c c c c c c c  }
\hline
\hline
Date (UT)& VLA Array & Band & Center Freq. & Bandwidth &HPBW$_{\rm maj}$& HPBW$_{\rm min}$&HPBW$_{\rm maj}$& HPBW$_{\rm min}$& PA\\
year/month/day& configuration&  & (GHz) & (GHz) & (\pr)  & (\pr) & (km) &(km) & ($^\circ$)  \\
\hline
2013/12/23 & B & Ku& 15.3& 5 & 0.38 & 0.32 & 1160 & 970 & --80 \\
2013/12/23 & B & Ku& 13.2& 1 & 0.46 & 0.36 & 1425 & 1115 & --82 \\
2013/12/23 & B & Ku& 14.2& 1 & 0.43 & 0.34 & 1335 & 1030 & --82 \\
2013/12/23 & B & Ku& 15.2& 1 & 0.40 & 0.32 & 1240 & 970 & --84 \\
2013/12/23 & B & Ku& 16.2& 1 & 0.38 & 0.30 & 1170 & 910 & --83 \\
2013/12/23 & B & Ku& 17.4& 1 & 0.34 & 0.32 & 1045 & 980 & --82 \\
2014/01/09 & B & Ku& 15.3& 5 & 0.40 & 0.30 & 1230 & 920 & --80 \\
2014/01/09 & B & Ku& 13.2& 1 & 0.46 & 0.36 & 1410 & 1120 & --86 \\
2014/01/09 & B & Ku& 14.2& 1 & 0.43 & 0.33 & 1315 & 1045 & --88 \\
2014/01/09 & B & Ku& 15.2& 1 & 0.44 & 0.31 & 1350 & 955 & --74 \\
2014/01/09 & B & Ku& 16.2& 1 & 0.41 & 0.29 & 1265 & 900 & --75 \\
2014/01/09 & B & Ku& 17.4& 1 & 0.39 & 0.28 & 1200 & 855 & --74 \\
2013/12/23 & B & X& 10.0& 4& 0.58 & 0.47 & 1775 & 1445 & --82 \\
2013/12/23 & B & X& 8.5& 1 & 0.70 & 0.55 & 2155 & 1700 & --83 \\
2013/12/23 & B & X& 9.5& 1 & 0.63 & 0.50 & 1920 & 1530 & --84 \\
2013/12/23 & B & X& 10.5& 1& 0.58 & 0.47 & 1770 & 1435 & --84 \\
2013/12/23 & B & X& 11.5& 1& 0.53 & 0.43 & 1625 & 1315 & --85 \\
2014/01/09 & B & X& 10.0& 5 & 0.61 & 0.48 & 1865 & 1480 & --72 \\
2014/01/09 & B & X& 8.5& 1 & 0.68 & 0.57 & 2095 & 1755 & --88 \\
2014/01/09 & B & X& 9.5& 1 & 0.61 & 0.51 & 1865 & 1580 & --84 \\
2014/01/09 & B & X& 10.5& 1 & 0.64 & 0.48 & 1975 & 1465 & --70 \\
2014/01/09 & B & X& 11.5& 1 & 0.58 & 0.44 & 1795 & 1345 & --70 \\
2014/05/04 & A & C & 6.0 & 4 & 0.78 & 0.78 & 2410 & 2410 & --45  \\
2014/05/04 & A & C & 4.5 & 1 & 0.78 & 0.78 & 2410 & 2410 & --45 \\
2014/05/04 & A & C & 5.5 & 1 & 0.78 & 0.78 & 2410 & 2410 & --45 \\
2014/05/04 & A & C & 6.5 & 1 & 0.78 & 0.78 & 2410 & 2410 & --45 \\
2014/05/04 & A & C & 7.5 & 1 & 0.78 & 0.78 & 2410 & 2410 & --45 \\
2014/05/04 & A & S & 3.0 & 2 & 0.63 &  0.6 &  1945 & 1840 & 73 \\
2014/08/16 & D & Ka&33.0& 8 & 2.5 & 2.2 & 7620 & 6735 & 54 \\
2014/08/16 & D & Ka&29.5& 1 & 3.0 & 2.7 & 9205 & 8315 & 63\\
2014/08/16 & D & Ka&30.5& 1 & 2.9 & 2.6 & 9030 & 7935 & 67\\
2014/08/16 & D & Ka&31.5& 1 & 2.8 & 2.4 & 8575 & 7275 & 56 \\
2014/08/16 & D & Ka&32.5& 1 & 2.7 & 2.3 & 8360 & 7090 & 54\\
2014/08/16 & D & Ka&33.5& 1 & 2.6 & 2.3 & 8135 & 6925 & 59\\
2014/08/16 & D & Ka&34.5& 1 & 2.6 & 2.2 & 7965 & 6670 & 57\\
2014/08/16 & D & Ka&35.5& 1 & 2.5 & 2.1 & 7635 & 6445 & 58\\
2014/08/16 & D & Ka&36.5& 1 & 2.4 & 2.0 & 7400 & 6250 & 56 \\
2014/12/27 & C & K & 22.0 & 8 & 1.2 & 0.8 & 3640 & 2490 & 44\\
2014/12/27 & C & K & 18.5 & 1 & 1.5 & 1.0 & 4555 & 2925 & 43\\
2014/12/27 & C & K & 19.5 & 1 & 1.4 & 0.9 & 4300 & 2770 & 42\\
2014/12/27 & C & K & 20.5 & 1 & 1.4 & 0.9 & 4190 & 2655 & 43\\
2014/12/27 & C & K & 21.5 & 1 & 1.3 & 0.8 & 3990 & 2530 & 43\\
2014/12/27 & C & K & 22.5 & 1 & 1.2 & 0.8 & 3815 & 2420 & 44\\
2014/12/27 & C & K & 23.5 & 1 & 1.2 & 0.7 & 3650 & 2305 & 44\\
2014/12/27 & C & K & 24.5 & 1 & 1.1 & 0.8 & 3355 & 2440 & 46\\
2014/12/27 & C & K & 25.5 & 1 & 1.1 & 0.8 & 3265 & 2325 & 45\\
\hline
\end{tabular}\label{tab.3}
\\

HPBW: Full beam width at half power in arcsec and in km on Jupiter, as used in the longitude-resolved maps. The HPBW has been taken at the equator near a longitude of 0$^\circ$. All values have been normalized to a geocentric distance of 4.24 AU, i.e. that for 23 December 2013. The position angle (PA) is given for the long (major) axis, counted clockwise from North. Note that these numbers deviate slightly from those listed in dP16. In particular at C band we degraded the resolution by a factor of close to 2.5 to enhance the signal-to-noise in the maps. 

One degree in latitude or longitude at Jupiter's disk center corresponds to $\sim$~1200 km at disk center.

\end{table*}

\end{document}